\documentclass[
reprint
,aps
,amssymb
,amsmath
,prx
,superscriptaddress
]{revtex4-1}

\usepackage{graphicx}
\usepackage{dcolumn}
\usepackage{bm}
\usepackage{color}
\usepackage[toc,page]{appendix}
\usepackage[colorlinks=true,allcolors=blue]{hyperref}
\usepackage[normalem]{ulem}

\newcommand{\mdc}[1]{{\color{black}#1}}
\newcommand{\mcm}[1]{{\color{black}#1}}

\begin{document}

\title{Hydrodynamics of shape-driven rigidity transitions in motile tissues}

\author{Michael Czajkowski}
\affiliation{Physics Department and Soft Matter Program, Syracuse University}
\author{Dapeng Bi}
\affiliation{Department of Physics, Northeastern University, Boston, MA 02115}
\author{M. Lisa Manning}
\affiliation{Physics Department and Soft Matter Program, Syracuse University}
\author{M. Cristina Marchetti}
\affiliation{Physics Department and Soft Matter Program, Syracuse University}
\date{\today}
 
\begin{abstract}
In biological tissues, it is now well-understood that mechanical cues are a powerful mechanism for pattern regulation.  While much work has focused on interactions between cells and external substrates, recent experiments suggest that cell polarization and motility might be governed by the internal shear stiffness of nearby tissue, deemed ``plithotaxis". Meanwhile, other work has demonstrated that there is a direct relationship between cell shapes and tissue shear modulus in confluent tissues. Joining these two ideas, we develop a hydrodynamic model that couples cell shape, and therefore tissue stiffness, to cell motility and polarization.  Using linear stability analysis and numerical simulations, we find that tissue behavior can be tuned between largely homogeneous states and patterned states such as asters, controlled by a composite ``morphotaxis" parameter that encapsulates the nature of the coupling between shape and polarization. The control parameter is in principle experimentally accessible, and depends both on whether a cell tends to move in the direction of lower or higher shear modulus, and whether sinks or sources of polarization tend to fluidize the system.

\end{abstract}

\pacs{64.70.D-, 87.18.Fx, 61.43.Er}
\maketitle


\section{Introduction}
Pattern formation during embryonic development, coordinated tissue movements in wound healing, and the breakdown of patterning in cancer tumorogenesis have all traditionally been explained in terms of biochemical signaling, such as morphogen gradients and growth factor secretion.  
Although biochemical gradients are clearly important, recent work has suggested that mechanical interactions and mechano-sensitive response can play a complementary and vital role in the robust patterning of these self-organized systems.  For example, the extra-cellular matrix (ECM) that contributes to the mechanical environment of cancer tissues strongly affects metastasis~\cite{Friedl2009,Paszek2005}, and the stiffness of an underlying substrate can control differentiation~\cite{Discher2005,Engler2006} and collective cell migration in wound healing assays for cell monolayers~\cite{Sunyer2016}.

Concurrent with these investigations of cell-substrate and cell-ECM interactions, another group of researchers has focused on cell-cell interactions, in an effort to understand the ``material properties" of tissues. Continuum models that describe epithelia as active viscoelastic fluids~\cite{Blanch-Mercader2017,Blanch-Mercader2017-2,Ranft2010,Ishihara2017,Yabunaka2017}  or active elastic sheets~\cite{Banerjee2012,Banerjee2011,Banerjee2015,Kopf2013} have been shown to reproduce many phenomena observed in wound healing assays and confined tissues.  Experimental studies discovered that many 2D monolayers~\cite{Angelini2011,Nnetu2012} and 3D bulk tissues~\cite{Schotz2008, Schotz2013, Pawlizak2015} are viscoelastic, exhibiting glassy dynamics that indicates they are close to a continuous fluid-to-solid, or jamming transition.  Developing continuum models that incorporate jamming transitions has proven difficult even in non-active materials~\cite{Falk2011,Sollich2006,Henann2013}, and so continuum models to date have not included this effect.
In addition, although most work has focused on the average material properties of a tissue, many tissues are heterogeneous. Therefore, given the close proximity of a fluid-solid transition where the shear modulus is expected to rise quickly from zero, it is natural to wonder if stiffness gradients within a tissue can drive patterning. There is already some experimental evidence for this; Tambe and coworkers coined the term ``plithotaxis" to describe their observation that MDCK cells polarize and move in the direction of local maximal principal stress to minimize local shear~\cite{Tambe2011}.

To our knowledge, there are no models that seek to quantify how gradients in stiffness within a tissue drive patterning, or predict the parameters that control patterning in such a system, although there are some analogues that can guide us.  For example, in active particle-based models, there is a direct relationship between the packing fraction of particles and the fluidity of the material. This leads to a natural coupling between the polarization (the direction that a particle wants to move) and the packing fraction that can be encapsulated in hydrodynamic models~\cite{Fily2012, Cates2010}
and gives rise to a novel type of patterning called motility induced phase separation.  Similarly, in liquid crystals there is a relationship between the nematic order parameter and the molecular mobility~\cite{Stark2003}. 
Again, one can write a hydrodynamic model that encapsulates this relationship and predicts pattern formation in liquid crystals.

But what is an appropriate hydrodynamic model for confluent tissues?  It is well-established that cells in a tissue can be polarized to move in a particular direction, suggesting that polarization should be a field in any hydrodynamic description, just as in flocking~\cite{Toner1998,Gowrishankar2016,Mishra2010} and particle-based active matter models.  But confluent tissues can change from fluid to solid at a packing fraction of unity, suggesting that density might not be an optimal choice for the hydrodynamic field. A recent body of work based on vertex models at the cellular scale suggests that confluent tissues exhibit a novel type of rigidity transition based on cell shape~\cite{Farhadifar2007,Staple2010,Bi2014,Bi2015,Bi2016,Park2015}. This body of work is based, however,  on a mesoscopic energy functional controlled by a single-cell parameter, namely the target shape index, which does not lend itself to a hydrodynamic description. 

\begin{figure}[!h]  
\begin{center}
    \includegraphics[width=0.48\textwidth]{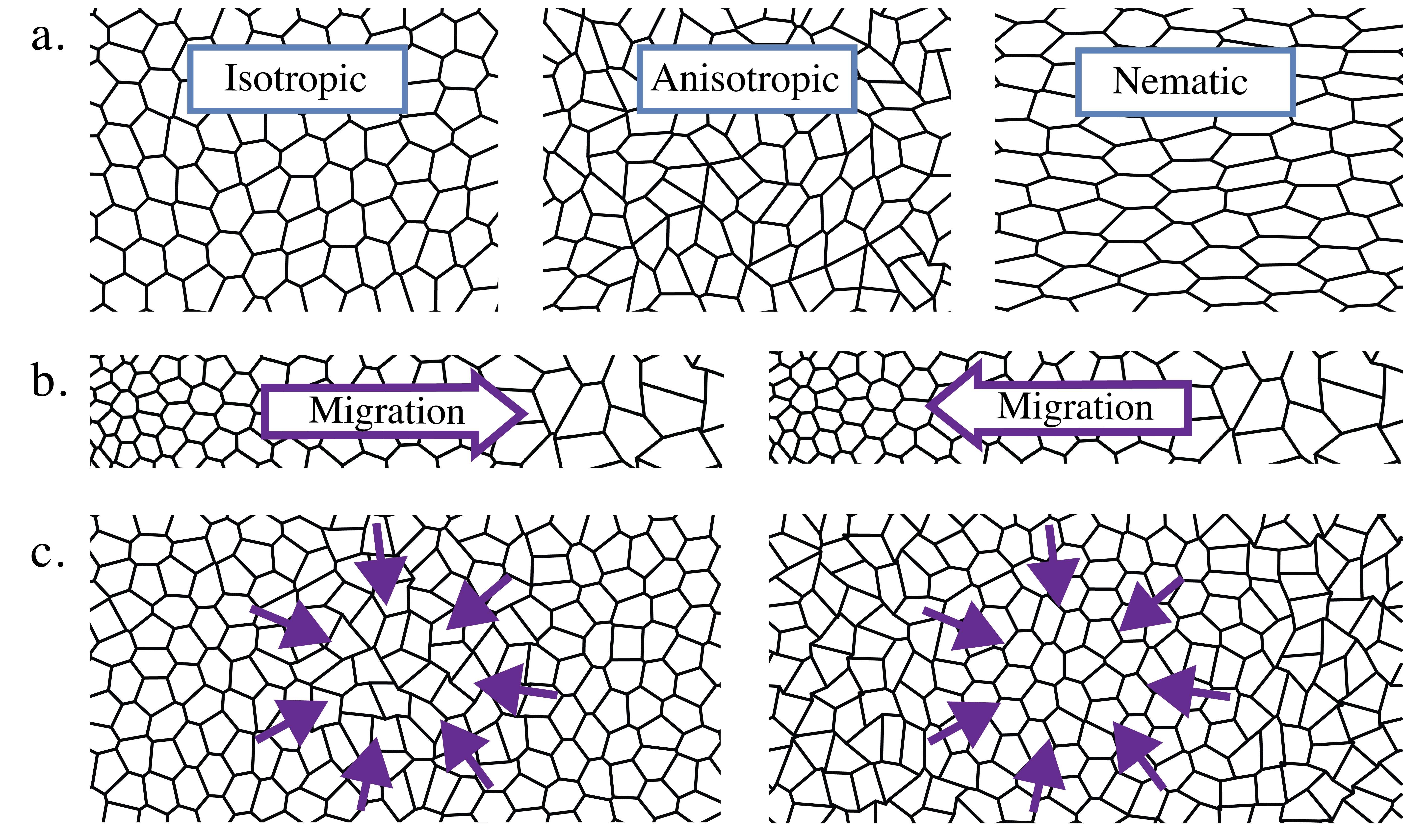}
    \caption{\label{fig:shapes_cartoon} Diagrams illustrating various shape-related behaviors in epithelial tissues. \textbf{a} From left to right: isotropic cell shapes (solid/jammed state), anisotropic cell shapes (fluid), nematic order of anisotropic cell shapes. \textbf{b} and \textbf{c} together display the \textit{morphotaxis} properties of the tissue. \textbf{b:} Cells may sense local gradients in shape, corresponding to gradients in tissue rigidity, and thereby polarize and migrate towards (left) or away from (right) the more anisotropic cells. \textbf{c:} Sinks of polarized motile forces may induce an increase (left) or a decrease (right) in the local  cell anisotropy.}
    \end{center}
\end{figure}

Therefore, in Section II of this manuscript, we develop a mean-field description of the fluid-solid transition in vertex models that directly incorporates our knowledge of how cell shapes govern jamming transitions and tissue stiffness in confluent tissues. It is important to note the distinction between a single-cell shape anisotropy field and an orientation field that captures alignment of elongated cells, first highlighted by Stark and Lubensky~\cite{Stark2003} for liquid crystals. In inert materials, however, molecular shape fluctuations decay on microscopic time scales and can therefore be neglected in hydrodynamic models.  Cells, in contrast,  are extended objects that can individually acquire isotropic or anisotropic shapes. Moreover, cellular shape changes have been shown to control the tissue rigidity, driving a continuous transition between liquid-like and solid-like states.  Shape fluctuations become long-lived at the transition and their dynamics must be incorporated in a hydrodynamic theory. When elongated, cells can additionally align their orientation and  form states with liquid crystalline order. Various shape-driven behavior of epithelial tissues are shown schematically in Fig.~(\ref{fig:shapes_cartoon}-a).
Recent work by Ishihara \textit{et al.}~\cite{Ishihara2017}, concurrent with our own, also uses vertex model energy for the tissue to construct a continuum theory. They focus on the case where cell elongation is always accompanied with nematic order of cellular orientation. In many tissues, however, one observes anisotropic cells without appreciable nematic order. For this reason, in this manuscript we neglect nematic order and simply consider the interplay of motility and shape changes.

In Section III, we incorporate the mean-field description of a shape anisotropy field into a hydrodynamic model with a coupling between cell shape anisotropy (e.g. tissue shear stiffness) and cell polarization. This introduces two important novel effects illustrated in Figs.~(\ref{fig:shapes_cartoon}-b,\ref{fig:shapes_cartoon}-c). The first is a ``plithotactic" parameter, which we take to be positive when cells migrate in the direction of stiffer (higher shear modulus) tissue, and negative when the cells migrate in the direction of softer tissue. 

The second effect captures how a sink of polarized motile forces affects tissue shape and shear stiffness. Our chosen convention is that if a sink (inward splay of polarization) tends to fluidize the tissue, generating anisotropic shapes (Fig.~\mbox{(\ref{fig:shapes_cartoon}-c)} left), the coupling parameter is negative, and positive in the opposite case (Fig.~\mbox{(\ref{fig:shapes_cartoon}-c)} right). As our analysis will demonstrate, these two effects encapsulate the interaction between polarization and shape and their product controls patterning.  Therefore we introduce the new term \textit{morphotaxis} -- morpho- from the greek $ \mu o\rho\phi\acute{\eta}$ meaning form or structure, and -taxis from the greek $\tau \acute{\alpha} \xi \iota \varsigma$. 
When the morphotaxis parameter is positive, patterns such as asters and traveling bands dominate. In contrast, when the morphotaxis parameter is negative, the tissue response is largely homogeneous. Finally, Section IV concludes with discussion of implications of this model for biological experiments and active matter more generally.

\begin{figure}[!t]  
\begin{center}
    \includegraphics[width=0.48\textwidth]{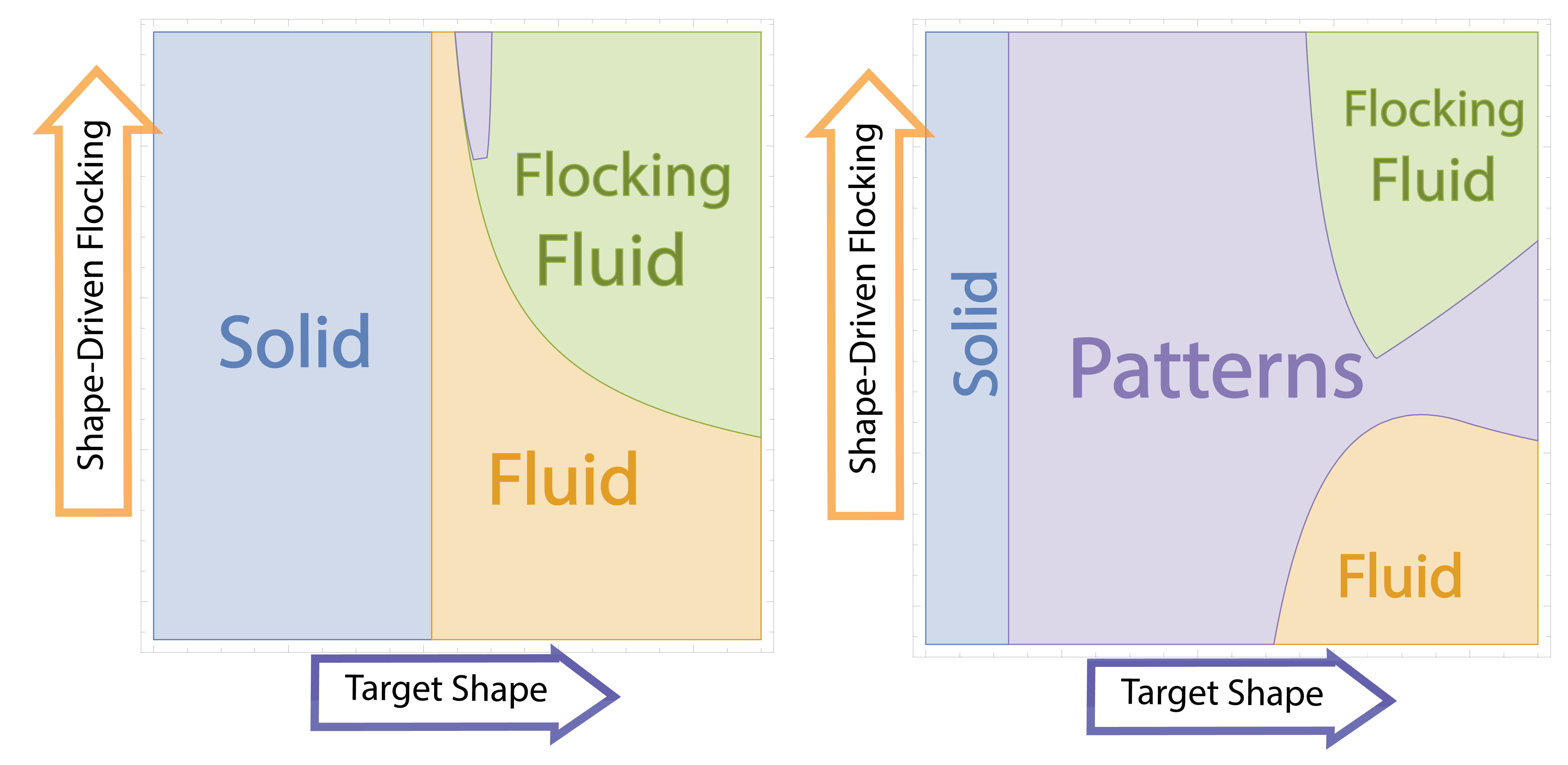}
    \caption{\label{fig:phases_cartoon} Schematic phase diagram comparing negative (left) and positive(right) \textit{morphotaxis} parameters for a shape-based hydrodynamic model where convergent polarization tends to decrease local cell shape anisotropy. The ``target shape" axis captures the average cell's preferred perimeter to area ratio, while the ``shape-driven flocking" axis quantifies the degree to which elongated cell shapes promote polarization.  The left panel corresponds to tissues in which cells tend to migrate toward fluid-like regions with more shape anisotropy, and the behavior is largely homogeneous. The right panel describes tissues where cells polarize toward solid-like regions of tissue with lower shape anisotropy, and the tissue exhibits patterns like asters or bands in a large region of the phase space.}
    \end{center}
\end{figure}

\section{A mean-field model for 2D shape anisotropy}
\label{sec:model}
\subsection{Review of rigidity in the Vertex Model}

The Vertex Model (VM) captures the topological features of confluent tissues by representing cells as polygons that tile the plane~\cite{Farhadifar2007,Hufnagel2007}. 
For a two-dimensional tissue containing $N$ cells the inter- and intra-cellular interactions are captured by a shape energy parametrized in terms of area $A_a$ and perimeter $P_a$  of the $a$-th cell, given by
\begin{equation}\label{eq:vertexenergy}
\mathrm{E}_{shape}=\sum_{a=1}^N\left[ \kappa_A (A_a-A_0)^2+\kappa_P (P_a-P_0)^2\right]\;.
\end{equation}
The first term arises from tissue incompressibility in three dimensions that allows cells to achieve a target area $A_0$ by adjusting their height. The second term captures the interplay between contractility of the actomyosin cortex and cell-cell adhesion, resulting in a cell membrane tension that controls the target perimeter $P_0$. $P_0$ increases with either decreasing cortical tension or with increasing cell-cell adhesion. Finally,
$\kappa_A$ and $\kappa_P$ are moduli associated with the area and perimeter terms, respectively.  

Numerical studies of the ground states of the shape energy given in Eq.~\eqref{eq:vertexenergy} have identified a rigidity transition that occurs as a function of  
the dimensionless ``target shape-index" $s_0 = P_0/\sqrt{A_0}$~ \cite{Bi2015,Bi2016,Park2015}. In previous work, the symbol $p_0$ was used for this quantity, but we change it here both for consistency with work in 3D~\cite{Merkel2017} and to distinguish it from cell polarization $p$. When $s_0<s_0^*\approx 3.81$, cortical tension dominates and the tissue is rigid with finite barriers to cellular rearrangements. For $s_0>s_0^*$ the energy barriers to cellular rearrangements vanish, resulting in zero-energy deformation modes that enable cells to elongate their shapes and fluidize the tissue. An analysis of cellular shapes reveals that the spatially-averaged cell shape-index $q=\left< P_a / \sqrt{A_a} \right>$ provides an order parameter for the transition in both non-motile and motile tissues: a tissue with $q< s_0^*$ is a rigid network of roughly regular cell shapes, while a tissue with $q>s_0^*$ is a fluid-like tissue of elongated and irregular cell shapes. 
\subsection{The Shape Tensor}
\label{subsec:shape_tensor}
Our first goal is to construct a continuum mean-field model of the rigidity transition captured by the VM. To do this we characterize the shape of the $a$-th cell via a shape tensor, given by
\begin{equation}\label{eq:shapetensor}
\mathbf{G}^a =\frac{1}{n_a} \sum_{\mu\in a}\left(\mathbf{x}_{\mu}-\mathbf{x}_{\,a}\right)\otimes\left(\frac{\mathbf{x}_{\mu}-\mathbf{x}_{\,a}}{|\mathbf{x}_{\mu}-\mathbf{x}_{\,a}|}\right) ,
\end{equation}
where $\mathbf{x}_\mu$ is the position of the $\mu$-th vertex of the $a$-th cell,
$\mathbf{x}_{\,a}$ points to the geometric center of cell-$a$ and the sum runs over the $n_a$ vertices on this cell.
\begin{figure}[!h]  
\begin{center}
    \includegraphics[width=0.48\textwidth]{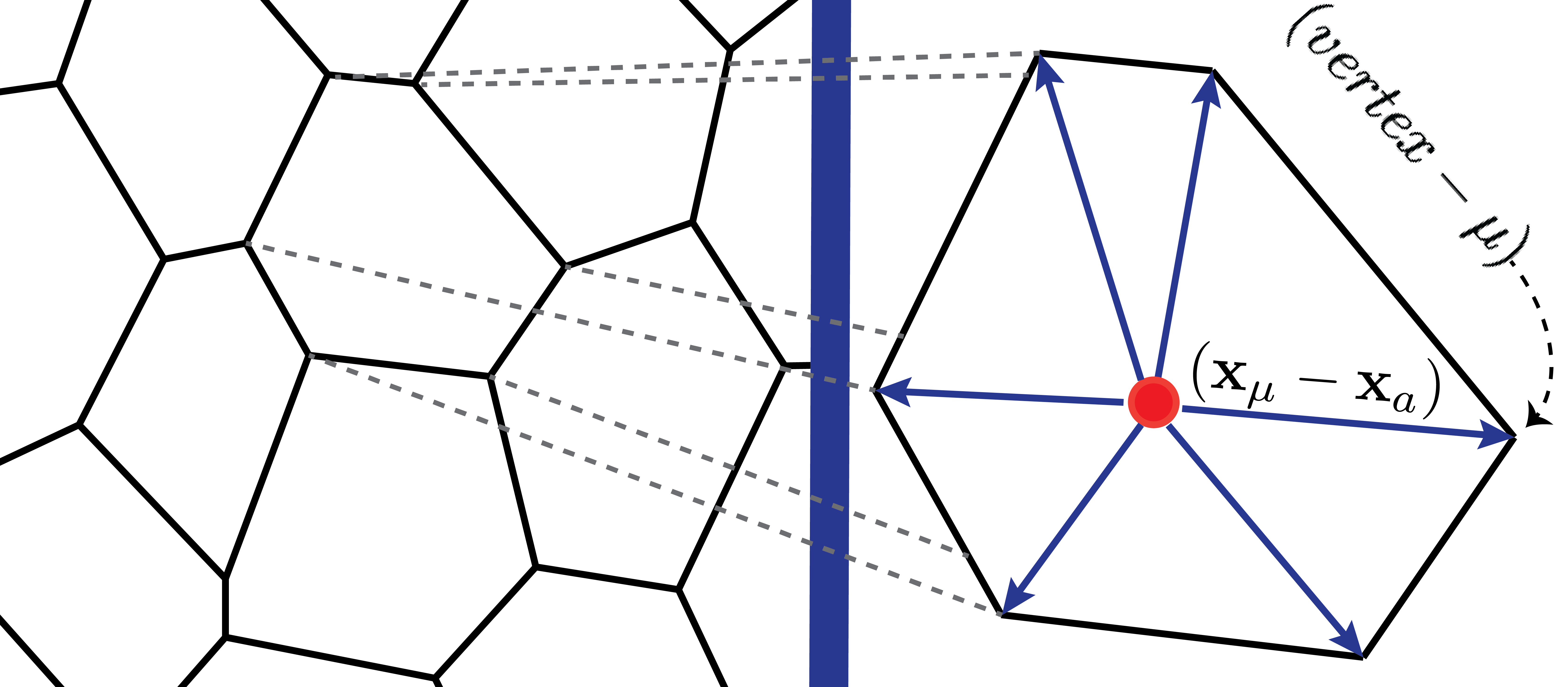}
    \caption{\label{fig:shape_tensor} Left: The Vertex Model representation of cells in a confluent monolayer. Right: The vectors which are used to create the cellular shape tensor.}
    \end{center}
\end{figure}
The cellular shape tensor $\mathbf{G}^a$  is very similar to the gyration tensor used to characterize the configuration of polymers \cite{Lee1997,Schroeder2005} -- our definition is slightly different because we want it to have units of length. Since $\mathbf{G}^a$ is a real and symmetric tensor, it has three independent degrees of freedom in two dimensions, and can generally be written in the form
\begin{equation}\label{eq:shapetensorgeneral}
G^a_{ij} =   M_a \left[\hat{e}^a_i \hat{e}_j^a - \frac12  \delta_{ij}  \right] +\frac12\Gamma_a\delta_{ij} \;,
\end{equation}
where $M_a=\lambda_1^a-\lambda_2^a>0$ and $\Gamma_a=\mathrm{Tr}[\mathbf{G}^a]=\lambda_1^a+\lambda_2^a$ are the sum and differences of the eigenvalues $\lambda_{1,2}^a$, $\mathbf{\hat{e}}^a$ is the eigenvector of the largest eigenvalue, $\lambda_1^a$, and $i,j$ denote Cartesian components. We introduce the dimensionless parameter $m_a=M_a/\Gamma_a$, which vanishes for isotropic cells and can be written as
\begin{eqnarray}
m_{a} =\frac{2}{\Gamma_a} \mathbf{\hat{e}}^a \cdot \mathbf{G}^a \cdot \mathbf{\hat{e}}^a - 1\;.
\label{eq:getm_tensor}
\end{eqnarray}
%
%
Note that $m_a$ is chosen to be positive definite. For regular $n$-sided polygons the shape tensor is always diagonal. Additionally, due to symmetry under rotations by $2\pi/n$, $\lambda_1^a=\lambda_2^a$, hence $m_a=0$. The area $A_a$ and the perimeter $P_a$ can then be expressed in terms of the shape tensor as
\begin{eqnarray}\label{eq:shapearea}
A_a &=& 2 n_a \sin\left(2 \pi /n_a\right) \mathrm{Det}[\mathbf{G}^a]\;,\\
\label{eq:shapeperi}
P_a &=& 2 n_a \sin\left( \pi /n_a\right) \mathrm{Tr}[\mathbf{G}^a]\;.
\end{eqnarray}

\subsection{\label{subsec:MFT}Mean-field theory}
Our first goal is to re-write the energy for a single cell (e.g. a single term in Eq.~\eqref{eq:vertexenergy}) for a regular polygon in terms of the cell shape anisotropy, $m_a$ using Eqs.~\eqref{eq:shapearea} and \eqref{eq:shapeperi}:
%
\begin{equation}
\begin{aligned}\label{eq:vertexenergyshape}
\epsilon_{a}=& \left[c_1(n_a)(1-m_a^2)\tilde{\Gamma}_a^2-1\right]^2 
+\tilde{\kappa} \left[c_2(n_a)\tilde{\Gamma}_a-s_0\right]^2 ,
\end{aligned}
\end{equation}
where $c_1(n_a)=\frac{n_a}{2}\sin(2 \pi/n_a)$, $c_2(n_a)=2 n_a \sin(\pi/n_a)$ and  we have scaled lengths with $\sqrt{A_0}$ and energies with  $A_0^2 \kappa_A$ and defined $\tilde{\Gamma}_a = \Gamma_a/\sqrt{A_0}$ and $\tilde{\kappa} = \kappa_P/(A_0 \kappa_A)$. 

Now we would like to use this to develop a simple mean-field model that captures the fluid-solid transition we see in metastable states at $s_0^* \approx 3.81$ in the vertex model. From previous work we expect the transition to be governed by the shape anisotropy $m_a$, so we minimize~\eqref{eq:vertexenergyshape} as function of $m_a$, keeping $\tilde\Gamma_a$ fixed such that $P_a = P_0$. Alternatively, we could have chosen to fix $\mathrm{Det}(\mathbf{G}_a)$ such that $A_a=A_0$, obtaining qualitatively the same results, as shown in Appendix \ref{app:shaperturb}.

\begin{figure}[!t]  
\begin{center}
    \includegraphics[width=0.5\textwidth]{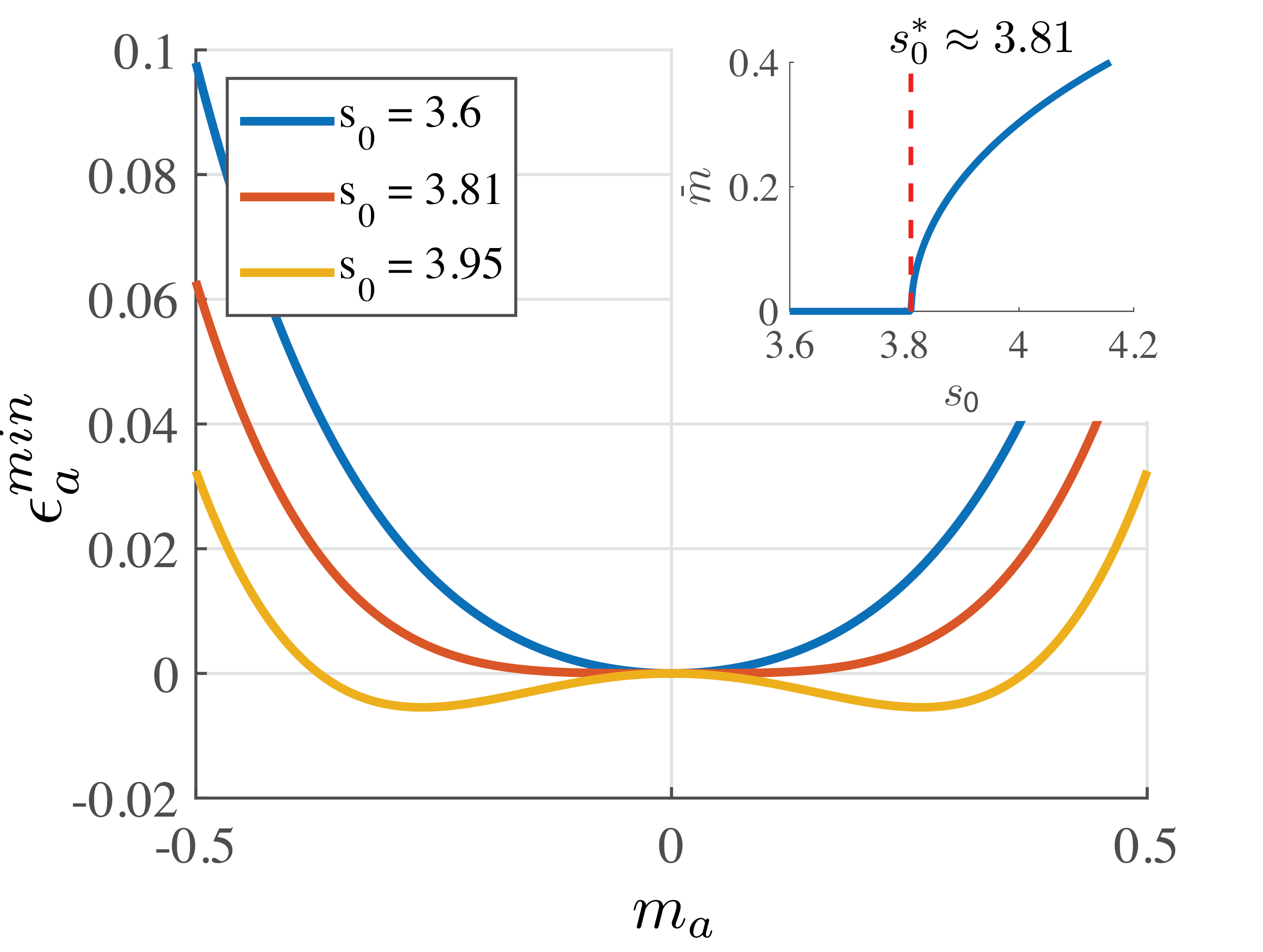}
    \caption{\label{fig:energy_doublewell} Mean-Field tissue energy as a function of shape-anisotropy for various values of the target shape-index $s_0$. As this shape index is increased past $s_0^* \approx 3.81$ the energy develops two minima and the anisotropy $\bar{m}$ becomes finite, as shown in the inset.}
    \end{center}
\end{figure}

The minimal single-cell energy can then be written as a function of cell shape anisotropy as
\begin{equation}\label{eq:freeenergyP}
\epsilon^{min}_a = \frac{1}{2}\alpha(s_0,n_a) m_{a}^2 + \frac{1}{4}\beta(s_0,n_a) m_{a}^4 \;.
\end{equation}
The parameters $\alpha$ and $\beta$ are controlled by the target shape parameter $s_0$ and the polygon degree $n_a$. While $\beta$ is positive for all $s_0$ and $n_a$, $\alpha$ changes sign as a function of $s_0$ and $n_a$. Equation~\eqref{eq:freeenergyP} has the familiar form of a $\phi^4$ theory, changing continuously from a single well to a double well at a critical value $s_0^*(n_a)$, as shown in Fig.~(\ref{fig:energy_doublewell}). 

Nothing in our analysis so far has specified $n_a$, the polygon degree, which sets the value of the shape order parameter at the critical point.
Previous work on the $2D$ vertex model has shown that the rigidity transition occurs at $s_0^*\approx 3.81$, which is the shape index corresponding to a regular pentagon. Although pentagons cannot tile space, we can still choose $n_a=5$ in our mean field model, so that the ground state anisotropy $\bar{m}$ that minimizes Eq.~\eqref{eq:freeenergyP} transitions from  $\bar{m}$ = 0 to $\bar{m}>0$ at the correct value of $s_0^*$, as shown in the inset to Fig.~(\ref{fig:energy_doublewell}). With this choice, $\alpha$ and $\beta$ are given by
\begin{eqnarray}\label{eq:alphatrace}
\alpha(s_0) &=& as_0^2 -bs_0^4 \;,\\
\label{eq:betatrace}
\beta(s_0) &=& bs_0^4 \; .
\end{eqnarray}
with $a=\mathrm{Cot}(\pi/5)/5$ and $b=\left[\mathrm{Cot}(\pi/5)\right]^2/100$. 
Cell-cell interactions could provide additional constraints not present in Eq.~\eqref{eq:freeenergyP}\mdc{, which should generally increase the energy of a cell (hence this is a \textit{minimal} energy)}. Recent work by some of us has also shown that in this model rigidity arises from purely geometric incompatibility~\cite{Merkel2017}, even in the absence of topological defects such as $T_1$ transitions~\cite{Moshe2017}. 

In summary, we have re-written the vertex model energy functional in terms of the shape anisotropy $m$ of regular polygons of degree $n$, minimized with respect to $m$ to find a ground state, and then chosen $n=5$ so that the ground state switches from isotropic to anisotropic shapes at a value of the control parameter that is consistent with simulations of the microscopic model.

%


\section{Hydrodynamic theory of cellular shape}
\label{sec:hydrodynamics}

Guided by the mean-field theory described in the previous section, we now formulate a continuum model of the shape-driven rigidity transition. As previously pointed out in the context of the Poisson-bracket derivation of the hydrodynamic equations of nematic liquid crystals \cite{Stark2003}, it is important to distinguish between  fluctuations in the shape of individual cells, as quantified by the single-cell anisotropy $m_a$, and fluctuations in the local alignment of elongated cells that are captured by correlations in the direction $\mathbf{\hat{e}}^a$ of the shape tensor eigenvector. 
To define continuum fields, it is convenient to introduce 
the traceless part of the cellular shape tensor, given by
\begin{equation}
\tilde{G}_{ij}^a= G_{ij}^a - \frac12 \delta_{ij} \Gamma_a = M_a\left[\hat{e}^a_i \hat{e}_j^a - \frac12  \delta_{ij}  \right]\;.
\label{eq:Gtilde}
\end{equation}
Following conventional definitions, we  introduce coarse-grained fields, given by
\begin{eqnarray}\label{eq:gammafield}
&&\Gamma(\mathbf{x}, t) = \left[ \sum_a \Gamma_a~ \delta(\mathbf{x}-\mathbf{x}_{a}) \right]_c \;,\\
\label{eq:Gfield}
&&\tilde{G}_{ij}(\vec{x},t)=\left[\sum_a \tilde{G}^a_{ij}~ \delta(\mathbf{x}-\mathbf{x}_{a})\right]_c\;,
\end{eqnarray}
where the  brackets $\left[...\right]_c$ denote coarse-graining and $\mathbf{x}_a$ is the position of the centroid of the $a$-th  polygonal cell.
Additionally,  the local coarse-grained number density is given by
\begin{equation}\label{eq:rhofield}
\rho(\mathbf{x}, t) = \left[ \sum_a \delta(\mathbf{x}-\mathbf{x}_{a}) \right]_c \;.
\end{equation}
For fixed number of cells, i.e., in the absence of cell growth and death, and in systems of fixed total area $A_T$ with periodic boundary conditions, the number density is slaved to cell area and does not fluctuate appreciably in confluent tissues. For this reason in the following we simply equate the density to its mean value $\rho_0=1/\overline{A}$, with $\overline{A}$ the mean cell area. 
The coarse-grained field $\Gamma(\mathbf{x},t)$ represents a fluctuating cell perimeter density. If all cell perimeters are identical it will simply be proportional to the number density.
The coarse-grained field $\tilde{G}_{ij}(\mathbf{x}, t)$ is a symmetric
and traceless tensor of rank two.
It has a structure similar to that of the familiar nematic alignment tensor, but it incorporates both
fluctuations in individual cell shape and in the direction of the principal eigenvector. To separately quantify cell-shape fluctuations, we  define an additional  coarse-grained field, the cell-shape anisotropy, as
\begin{equation}\label{eq:mfield}
m(\mathbf{x}, t) = \frac{\left[ \sum_a M_a\delta(\mathbf{x}-\mathbf{x}_{a}) \right]_c}{\Gamma(\mathbf{x}, t)} \;.
\end{equation}
The traceless shape tensor is then written as
\begin{equation}
\tilde{G}_{ij}(\mathbf{x},t)=m(\mathbf{x},t)\Gamma(\mathbf{x},t)Q_{ij}(\mathbf{x},t)\;,
\label{eq:Gcont}
\end{equation}
where
\begin{equation}\label{eq:Qfield_define}
Q_{ij}(\mathbf{x}, t) = \frac{\tilde{G}_{ij}(\mathbf{x},t)}{m(\mathbf{x}, t)\Gamma(\mathbf{x}, t)} \;.
\end{equation}
is the nematic alignment tensor. 

At the single-cell level, the shape tensor $G_{ij}^a$ is characterized by three independent quantities that can be chosen as the cell area (proportional to $\mathrm{Det}(\mathbf{G}_a)$ and inversely proportional to the mean density in a confluent tissue), the cell anisotropy $m_a=M_a/\Gamma_a$, and the angle defined by $\mathbf{\hat{e}}_a$. Then $\Gamma_a$, which is proportional to cell perimeter, can be written as 
$\Gamma_a=c\sqrt{A_a/(1-m_a^2)}\sim \left[\rho_0(1-m_a^2)\right]^{-1/2}$, with $c$ a numerical constant of order unity. Fluctuations in the field $\Gamma(\mathbf{x},t)$ will then be controlled by density and shape anisotropy fluctuations, and
$\Gamma(\mathbf{x},t)=\Gamma\left(\rho(\mathbf{x},t),m(\mathbf{x},t)\right)\simeq \Gamma\left(\rho_0,m(\mathbf{x},t)\right)$. In other words, we do not need to consider $\Gamma$ as an independent field as it is slaved to $m$.

If cells are isotropic, both $m$ and $\tilde{G}_{ij}$ vanish identically. When cells are elongated and $m$ is finite, cells can additionally exhibit orientational order captured by the tensor $Q_{ij}$. 
For uniaxial systems, $Q_{ij}$ can be written as
\begin{equation}
Q_{ij}(\mathbf{x},t)=S(\mathbf{x},t)\left[n_in_j-\frac12\delta_{ij}\right]\;,
\label{eq:Qfield}
\end{equation}
where $\mathbf{n}(\mathbf{x},t)$ is the nematic director.  Tissues of elongated cells with a nonzero mean value of $m$ can then additionally exhibit orientational order of cell elongation characterized by a finite value of  $S(\mathbf{x},t)$.
Such nematic order has not, however, been observed in simulations of Active Vertex or Self-Propelled Voronoi models in the absence of interactions that tend to align cell polarization. For this reason we do not consider  the dynamics of $Q_{ij}$ here and leave this for future work. As seen below, here we only model tissues where cell elongation may result in polar alignment of cell motility, possibly leading to global flocking of the tissue. This may describe monolayers of MDCK cells as studied in Ref. ~\cite{Puliafito2012} that show a strong correlation between cell morphology and the transition between motile and non-motile tissues.

\subsection{Hydrodynamics of Shape in Non-Motile Tissues}

We begin by constructing a  hydrodynamic equation for $m(\mathbf{x},t)$ in the absence of cell motility. Due to the complexity of the interactions arising from the shape energy, an exact coarse graining appears intractable. Instead, we recognize that the simplified mean-field theory of pentagons described in Section \ref{subsec:MFT} already encodes the key properties of the shape driven liquid-solid transition seen in simulations ~\cite{Bi2015,Bi2016}. At large length scales, we then neglect density fluctuations and assume that the VM can be described by a Landau-type free energy functional given by 
%
\begin{equation}\label{eq:mfieldenergy}
\mathrm{F} =  \int \mathrm{d}\mathbf{x} \left\{ \frac{1}{2}\alpha(s_0)m^2 + \frac{1}{4}\beta(s_0)m^4 + \frac{\mathrm{D}}{2} (\bm{\nabla} m)^2 \right\}\;,
\end{equation}
where $\mathrm{D}$ is a stiffness that describes the energy cost of spatial variation in cellular shape arising from interactions. 
Since the rigidity transition is found to be continuous in numerical simulations of Vertex and Voronoi models, and well described by the free energy of Eq.~\eqref{eq:mfieldenergy}, we use here the same quadratic energy derived for a single cell as a mean-field description for the tissue. The relaxational dynamics of $m(\mathbf{x},t)$ is then given by
\begin{equation}
\begin{aligned}\label{eq:mfieldstatic}
\partial_t m= &-\frac{1}{\gamma} \frac{\delta F}{\delta m} \\
 = &-\left[\alpha(s_0)+\beta(s_0) m^2\right]m + \mathrm{D}\nabla^2m\;,
\end{aligned}
\end{equation}
where for simplicity we have taken the kinetic coefficient $\gamma=1$. The phenomenological parameters $\alpha$ and $\beta$ depend on the target shape index $s_0$ via Eqs.(\ref{eq:alphatrace}, \ref{eq:betatrace}), with $\beta>0$ and $\alpha$ changing sign at $s_0=3.81$. The steady state solution of Eq.~\eqref{eq:mfieldstatic} then yields two homogeneous states: a solid state with $m_{ss}=0$ for $\alpha>0$, corresponding to $s_0<3.81$, and a liquid state with $m_{ss} = \sqrt{-\alpha/\beta}$ for $\alpha<0$, corresponding to $s_0>3.81$. It therefore provides a mean-field description of the liquid-solid transition seen in the vertex model. The stiffness $\mathrm{D}$ tends to stabilize the homogeneous states. Fluctuations are characterized by a correlation length $\ell_m \sim\sqrt{\mathrm{D}/|\alpha|}$ that diverges at the transition. 
In the rest of this work $\alpha$ and $\beta$ are functions of $s_0$ even where this dependence is suppressed.



\subsection{Hydrodynamics of Shape in Motile Tissues}



Inspired by the Toner-Tu model of flocking,  we describe cell motility at the continuum level in terms of a local polarization field, $\mathbf{p}(\mathbf{x},t)$, that defines the direction of  the propulsive force originating from the traction that cells exert on a substrate. 
A non-zero value of $|\mathbf{p}|$ describes the situation where cells align their direction of polarization, exerting a coordinated thrust in a common direction that spontaneously breaks rotational symmetry. In particle-based flocking models, a mean polarization arises from the explicit tendency of particles to align with their metric neighbors and is thereby tuned by density. In contrast, collective motion in our model is directly tuned by cell shape, 
which can exhibit slow dynamics at the liquid-solid transition.
Neglecting for now the possibility of nematic order of elongated cell shapes, the large scale, long time dynamics of the tissue is then described by coupled continuum equations for cell anisotropy and polarization, given by
\begin{widetext}
\begin{equation}\label{eq:mdot}
\partial_t m+\nu_1\mathbf{p}\cdot\mathbf{\nabla}m=-\left[\alpha(s_0)+\beta(s_0) m^2\right]m + \sigma\mathbf{\nabla}\cdot\mathbf{p}+\mathrm{D}\nabla^2m\;,
\end{equation}
\begin{equation} \label{eq:pdot}
\partial_t\mathbf{p}+\lambda_1\left(\mathbf{p}\cdot\mathbf{\nabla}\right)\mathbf{p}=-\left[\alpha_p(m) + \beta_p p^2\right]\mathbf{p}-\nu\mathbf{\nabla}m+\lambda_2\mathbf{\nabla}p^2-\lambda_3\left(\mathbf{\nabla}\cdot\mathbf{p}\right)\mathbf{p}+\mathrm{D}_p\nabla^2\mathbf{p}\;.
\end{equation}
\end{widetext}
As with all phenomenological hydrodynamic models, Eqs.~\eqref{eq:mdot} and \eqref{eq:pdot} contain quite a few parameters, which can in general be functions of $m$ and $p^2$. For simplicity here we take them as constant unless otherwise noted.  The cell anisotropy field $m$ is convected by polarization at rate $\nu_1$ and diffuses with diffusivity $\mathrm{D}$. The polarization equation has a form closely analogue to the Toner-Tu equations, with the shape anisotropy $m$ replacing the density, but with the important difference that $m$ is not conserved. The convective parameters $\lambda_1, \lambda_2$ and $\lambda_3$ arise from the breaking of Galilean invariance due to the presence of the substrate. For simplicity we neglect the anisotropy of the stiffnesses for bend and splay deformations and assume a single isotropic diffusivity, $\mathrm{D}_p$. The coefficients $\beta$ (described in Section II) and $\beta_p$ are both assumed to be positive so the model admits stable anisotropic and flocking states.
Both $\alpha$ (introduced in the previous section) and $\alpha_p(m)=\alpha_p^0 - a m$ (with $\alpha_p^0,a>0$) change sign as a function of $s_0$, resulting in mean-field transitions and instabilities tuned by the target cell shape $s_0$.
The choice $a>0$ describes the possibility that anisotropic cell shapes promote flocking in the fluid, which is a new ingredient of our model.
Since $a$ controls the onset of flocking and its value is not experimentally constrained, we explore the stability of the hydrodynamic model as a function of this parameter.

There are two key parameters that couple $\mathbf{p}$ and $m$. The term proportional to $\sigma$ describes the fact that spatial gradients of polarization can drive changes in local cell shape. A positive value of $\sigma$ corresponds to a situation where $m$ increases towards regions of positive polarization splay. The sign of this parameter could be determined by correlating TFM measurements of local traction forces with cell shape fluctuations from segmentation images of static tissues. Here we set $\sigma = +1$. 
The term proportional to $\nu$ represents a pressure gradient driven by cellular shape. Following Ref.~\cite{Tambe2011}, we will refer to $\nu$ as the \emph{plithotactic} parameter because 
its sign controls whether cells prefer to migrate towards stiffer solid-like regions of the tissue ($\nu>0$) or towards soft fluid-like ones ($\nu<0$). As discussed in the introduction, wound healing assays in expanding tissues have reported the tendency of MDCK cells to migrate along directions of minimal shear stresses, which would suggest a tendency to move from the solid to the liquid, corresponding to $\nu<0$~\cite{Tambe2011}, although other behavior may occur in different cell types. Therefore, we explore the hydrodynamic model for $\nu = +1$ and for $\nu =-1$.


An important difference between the Toner-Tu equations and our model is that cell-shape anisotropy $m$ is not a conserved field, but an order parameter associated with a liquid solid transition. Our model couples for the first time collective cell motility with a tissue rigidity transition, allowing us to examine the feedback between motility and shape in a crowded environment. 

\subsection{Homogeneous Steady States}
Our hydrodynamic equations for motile tissues exhibit three homogeneous steady state solutions: 

(i) a solid  with $m_{ss}=p_{ss}=0$ for $\alpha>0$ and $\alpha_p^0>0$, corresponding to a non-motile rigid tissue with isotropic cellular shapes; 

(ii) a non-motile fluid with $m_{ss}=\sqrt{-\alpha/\beta}$ and $p_{ss}=0$  for $\alpha<0$ and $\alpha_p(m_{ss})>0$, or equivalently $-\beta(\alpha_p^0/a)^2<\alpha<0$,  corresponding to a liquid-like tissue with elongated  cellular shapes and zero mean motion; \\ and

(iii) a flocking fluid with $m_{ss}=\sqrt{-\frac{\alpha}{\beta}}$ and $p_{ss} =\sqrt{( a m_{ss} -\alpha_p^0)/\beta_p}$ for $\alpha<0$ and $\alpha_p(m_{ss})<0$, or equivalently $\alpha<-\beta(\alpha_p^0/a)^2$, corresponding to a liquid-like tissue with elongated  cellular shapes and finite mean polarization. 

  The regions of parameter space where each solution exists are summarized in Table~(\ref{tab:steadystate}) and in Fig.~(\ref{fig:phases_sims_combined}). We find two critical values of $\alpha(s_0)$ in the mean-field phase diagram, corresponding to  $\alpha_{c1}=0$ and $\alpha_{c2}=-\beta(\alpha_p^0/a)^2$.  These give two critical lines in the $(s_0,a)$ phase diagram shown in Fig.(\ref{fig:phases_sims_combined}), where $s_0$  is the target shape parameter  and $a$ controls elongation-driven collective motility.

\begin{table*}[!t]
\begin{centering}
\bigskip
\begin{tabular}{|c|c|c|}
\hline
\rule[-0.5ex]{0pt}{3.5ex} Phase & Fields & Homogeneous Stability Condition \\ \hline \hline
\rule[-0.5ex]{0pt}{3.5ex} Solid & $m_{ss}=|p_{ss}|=0$ & $\alpha>0$ , $\alpha_p^0>0$ \\\hline
\rule[-0.5ex]{0pt}{3.5ex} Fluid & $m_{ss}^2=-\frac{\alpha}{\beta}$ \, ,\, $|p_{ss}|=0$ & $\alpha<0$ , $\alpha_p(m_{ss}) >0$ \\\hline
\rule[-0.5ex]{0pt}{3.5ex} Flocking Fluid \; & \; $m_{ss}^2=-\frac{\alpha}{\beta}$ \, , \, $\beta_p p_{ss}^2 = -\alpha_p(m_{ss})$ \; &  $\alpha<0$, $\alpha_p(m_{ss})< 0$ \\
\hline
\end{tabular}
\bigskip
\end{centering}
\caption{Homogeneous Steady States. \label{tab:steadystate}} 
\end{table*}
Our model  yields a density-independent flocking transition in confluent tissues tuned by cortical tension and cell-cell adhesion, which are captured by 
the parameter $s_0$. 
The existence of a ``flocking solid" state has been prevented by the choice $\alpha_p^0>0$.

Our hydrodynamic equations are formally similar to those studied by Yang \textit{et al.} \cite{Yang2014} to describe populations of self-propelled entities in the absence of number conservation,  with a nonconserved density taking the place of the shape parameter $m$. This work, in fact, reports static and dynamical patterns qualitatively similar to the ones obtained here. One difference, however, is that the density of self-propelled entities discussed in Ref.~\cite{Yang2014} even if not conserved always fluctuates around a finite value, so that small fluctuations can have either sign. Here, the shape parameter $m$ is defined positive and fluctuations in the solid state where $m_{ss}=0$ can only be positive, describing the occurrence of liquid-like regions in a solid matrix.
This impacts the linear stability of these states, as discussed in the next section.

\subsection{\label{subsec:linstab}Linear Stability Analysis}

Here we examine the linear stability of each of the three homogeneous states against spontaneous fluctuations. After linearizing the hydrodynamic equations \eqref{eq:mdot} and \eqref{eq:pdot} in the fluctuations of the fields around their steady state values, 
$\delta m(\mathbf{x},t)=m(\mathbf{x},t)-m_{ss}$ and $\delta \mathbf{p}(\mathbf{x},t)=\mathbf{p}(\mathbf{x},t)-\mathbf{p}_{ss}$, we expand the fluctuations in Fourier components,  
\begin{equation}\label{eq:perturbations}
\left[ \begin{array}{c} \delta m(\mathbf{x},t) \\ \delta \mathbf{p}(\mathbf{x},t)  \end{array} \right] = \int\mathrm{d}\mathbf{k}~e^{ -i \mathbf{k} \cdot \mathbf{x}} \left[ \begin{array}{c} m_{\mathbf{k}}(t) \\  \mathbf{p}_{\mathbf{k}}(t) \end{array} \right]\;.
\end{equation}
%
The linear dynamics of the Fourier components of the fluctuations can then be written in the compact form
\begin{equation}
\label{eq:Mmatrix}
\partial_t\mathbf{\phi}_{\mathbf{k}}(t)=\mathbf{M}^{ss}(\mathbf{k})\cdot\bm\phi_{\mathbf{k}}(t)\;,
\end{equation}
where $\bm\phi_{\mathbf{k}}=\left(m_{\mathbf{k}},\mathbf{p}_{\mathbf{k}}\right)$ and $\mathbf{M}^{ss}(\mathbf{k})$ is a matrix given in Eqs.~(\ref{eq:matrix1}) and (\ref{eq:matrix3}) of Appendix \ref{app:linstab}.

\begin{figure*}[t]  
\begin{centering}
    \includegraphics[width=0.95\textwidth]{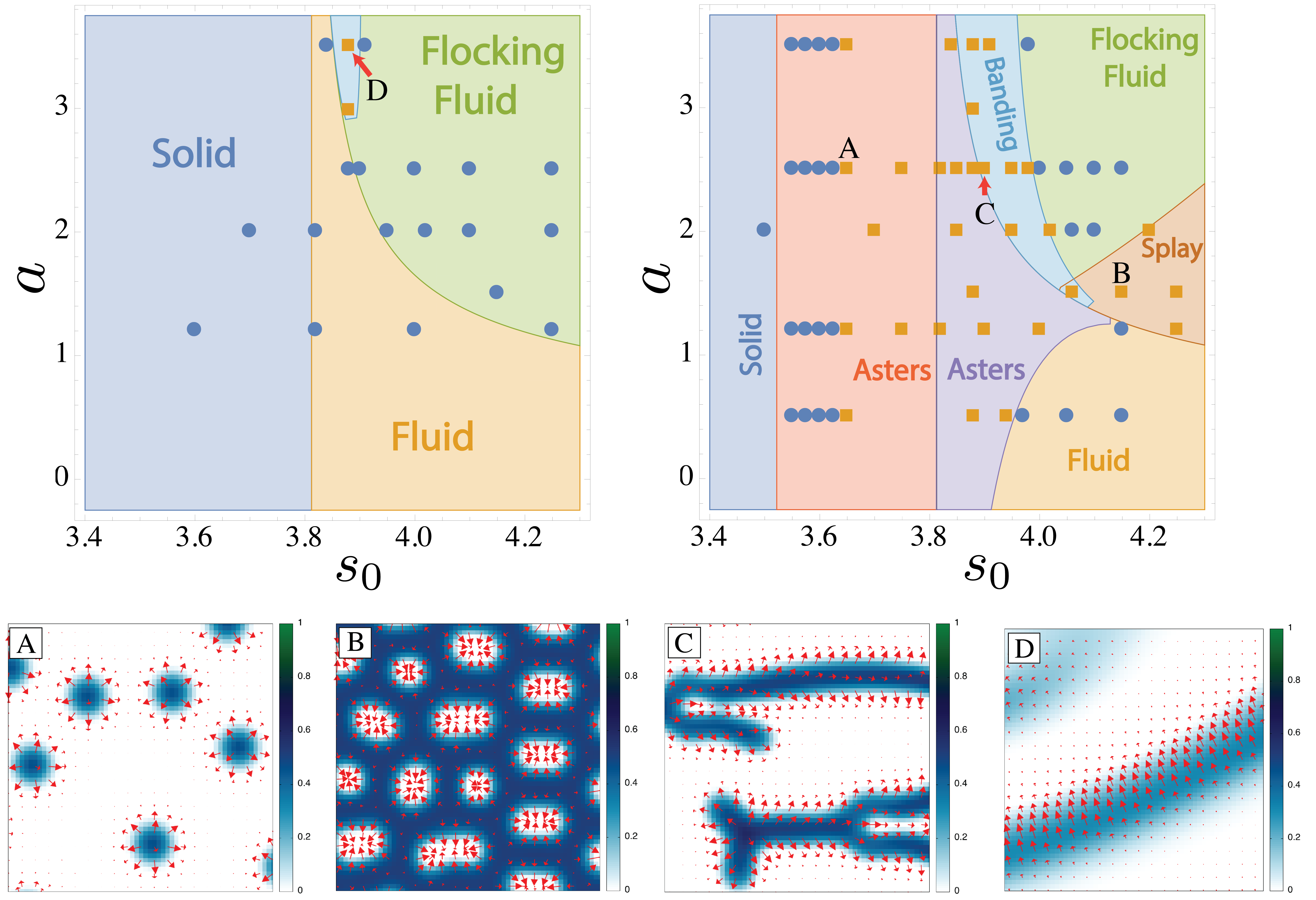}
    \caption{\label{fig:phases_sims_combined} Phase diagrams and simulation results in the $s_0-a$ plane. Blue circles represent simulations in which the fields relax to their homogeneous steady state solution. Orange squares represent simulations in which patterns are found to emerge. Here we compare the cases $\nu=-1$ (Top-Left) and $\nu=1$ (Top-Right) to show the qualitative change induced by this plithotactic parameter. Bottom: Snapshots of different types of emergent patterns from tissue simulations. Colorbars represent the magnitude of local anisotropy $(m)$ while red arrows represent local cell polarization $(\mathbf{p})$. (A): Sparse aster-like islands of anisotropic cells emerge near the onset of instability in the solid phase. (B): shows an example of ``solid'' islands arising in the flocking fluid phase due to a splay instability and preventing collective motion. (C) shows the elongated structures resultant from banding instability for $\nu=1$ while (D) shows the qualitatively different band structures for $\nu=-1$.}
\end{centering}
\end{figure*}

The decay or growth of the fluctuations is governed by the eigenvalues $z_\mu(\mathbf{k})$ of $\mathbf{M}^{ss}(\mathbf{k})$, where $\mu$ labels the eigenvalue (see Appendix \ref{app:linstab} for details).  
An instability occurs when ${\rm Re}[z_\mu(\mathbf{k})]>0$ for any $(\mu,\mathbf{k})$. A nonzero imaginary part of the eigenvalue corresponds to propagating modes. 

As we will see below, 
pattern formation in our model depends crucially on the sign of the product $\sigma\nu$ that defines the morphotaxis parameter of the tissue (or, since we have chosen $\sigma=+1$, the sign of $\nu$) and is best discussed by examining each steady state one at a time. This product combines the response of polarization to gradients in shape with the response of shape to sinks/sources of polarization.

\paragraph{Solid State.} The solid state with $m_{ss}=p_{ss}=0$ exists for $\alpha>0$. The steady state has no spontaneously broken symmetry  and fluctuations are isotropic in the sense that their decay rates only depend on the magnitude of $\mathbf{k}$, not on its direction. In this case it is convenient to split $\mathbf{p}_{\mathbf{k}}$ in components longitudinal and transverse to $\mathbf{k}$ as
$\mathbf{p}_{\mathbf{k}}=\left({p}^L_{\mathbf{k}}, {p}^T_{\mathbf{k}}\right)$, where ${p}^L_{\mathbf{k}}=\mathbf{\hat{k}}\cdot\mathbf{p}_{\mathbf{k}}$ and $\mathbf{p}^{\,T}_{\mathbf{k}}=\mathbf{p}_{\mathbf{k}}-\mathbf{\hat{k}}{p}^L_{\mathbf{k}}$, 
with $\mathbf{\hat{k}}=  \mathbf{k}/|\mathbf{k}|$. Fluctuations in the transverse part of the polarization that corresponds to bend deformations are decoupled and always decay. The coupled dynamics of fluctuations in shape anisotropy and ${p}^L_{\mathbf{k}}$ that describes splay deformation is controlled by two eigenvalues, given by
\begin{equation} 
\begin{aligned}\label{eq:eigenvals1}
z^{(solid)}_{\pm}=&-\frac12\left[\alpha+\alpha_p^0+(\mathrm{D}+\mathrm{D}_p)k^2\right] \\ 
& \pm \frac12\sqrt{\left[\alpha-\alpha_p^0+(\mathrm{D}-\mathrm{D}_p)k^2\right]^2+4 k^2 \nu \sigma}\;.
\end{aligned}
\end{equation}
%
%
The modes are always stable for $\sigma\nu<0$. When $\sigma \nu>0$ the mode $z^{(solid)}_+$ can become positive and yield an instability when  $\sigma \nu > \left[\sqrt{\alpha_p \mathrm{D}} + \sqrt{\alpha \mathrm{D}_p}\right]^2$. This condition is, however, obtained by relinquishing the constraint that $m>$ and allowing it to fluctuate freely around $m_{ss}=0$. Imposing the constraint of positive $m$ renormalizes the stability boundary.  Lacking an analytic tool, the analysis must, however, be carried out numerically.

The wavelength of the fastest growing mode defines a characteristic length scale given by
\begin{equation}\label{eq:solidlength}
\ell_{solid} =  2 \pi \sqrt{\frac{2 \mathrm{D} \mathrm{D}_p}{\sigma \nu - \alpha_p^0 \mathrm{D} - \alpha \mathrm{D}_p}} \; .
\end{equation}
At the onset of instability this becomes  $\ell_{solid}=2\pi(\mathrm{D}\mathrm{D}_p/\alpha\alpha_p^0)^{1/4}$ and can be interpreted as the geometric mean of two length scales, $\ell_{solid} =2\pi\sqrt{\ell_m \ell_p}$, where $\ell_m = \sqrt{\mathrm{D}/\alpha}$ represents the distance over which diffusion balances the relaxation of the anisotropy field, while $\ell_p = \sqrt{\mathrm{D}_p/\alpha_p^0}$ describes spatial variation in the polarization field.


\paragraph{Fluid state.} The non-polarized fluid state is obtained for  $\alpha<0$ and $\alpha_p=\alpha_p(m_{ss})>0$ and  has finite $m_{ss}=\sqrt{-\alpha/\beta}$ and $p_{ss}=0$.  The behavior is formally the same as obtained for the solid state, but with the relaxation rate of the anisotropy parameter $m$ replaced by $-2\alpha>0$ and that of  polarization decreased from $\alpha_p^0$ to $\alpha_p=\alpha_p^0-am_{ss}=\alpha_p^0-a\sqrt{-\alpha/\beta}>0$. The steady state is again isotropic 
and fluctuations in the transverse polarization $\vec{p}^T_{\mathbf{k}}$ are decoupled and always decaying. The coupled dynamics of fluctuations in shape and splay polarization is controlled by the eigenvalues
\begin{equation} 
\begin{aligned}\label{eq:eigenvals2}
z^{(fluid)}_{\pm}=&-\frac12\left[2 |\alpha|+\alpha_p+(\mathrm{D}+\mathrm{D}_p)k^2\right] \\ 
& \pm \frac12\sqrt{\left[2|\alpha|-\alpha_p+(\mathrm{D}-\mathrm{D}_p)k^2\right]^2+4 k^2 \nu \sigma}\;.
\end{aligned}
\end{equation}
Again the steady state is stable when $\sigma \nu <0$ and unstable for $\sigma \nu> \left[\sqrt{\alpha_p \mathrm{D}} + \sqrt{2 |\alpha| \mathrm{D}_p}\right]^2$. The wavelength of the fastest growing mode is
\begin{equation}\label{eq:fluidlength}
\ell_{fluid} =  2 \pi \sqrt{\frac{2 \mathrm{D} \mathrm{D}_p}{\sigma \nu - \alpha_p \mathrm{D} - 2 |\alpha| \mathrm{D}_p}} \; 
\end{equation}
that reduces to $\ell_{fluid} = 2\pi(DD_p/2|\alpha|\alpha_p)^{1/4}$ at the onset of the instability. Note, however, that $\alpha_p$ vanishes at $|\alpha_{c2}|=(\alpha_p^0/a)^2\beta$ where the system undergoes a mean-field transition to a flocking liquid state and $\ell_{fluid}$ diverges.

\paragraph{Flocking fluid.} In the flocking fluid state, obtained for  $\alpha<\alpha_{c2}$,   the system acquires a finite mean polarization, breaking rotational symmetry, and all modes are coupled.  We  then choose the $x$ axis  along the direction of broken symmetry, i.e., $\mathbf{p}_{ss} = p_{ss} \mathbf{\hat{x}}$. For simplicity we only examine here the behavior of the fluctuations for wavevectors parallel and perpendicular to the direction of broken symmetry. For
wavevector $\mathbf{k}$ along the direction of broken symmetry, $\mathbf{k} = k \mathbf{\hat{x}}$, \textit{bending} fluctuations in the orientation of polarization, $\delta p^y_{\mathbf{k}}$, decouple and are always stable. Fluctuations in shape anisotropy and  the magnitude of polarization,  $\delta p^x_{\mathbf{k}}$, are coupled and the  stability is controlled by the eigenvalues
\begin{widetext}
\begin{equation}
\begin{aligned}\label{eq:bandingrates}
2 z^{(band)}_{\pm}&= 2(\alpha +\alpha_p(m_{ss})) +\textit{i} p_{ss} (\nu_1 + \lambda_{T}) k - (\mathrm{D} +\mathrm{D}_p) k^2 \\
 \pm& \sqrt{\left[2(\alpha-\alpha_p(m_{ss}))+\textit{i} p_{ss} (\nu_1-\lambda_{\mcm{T}}) k - (\mathrm{D}-\mathrm{D}_p) k^2\right]^2+4\sigma(\nu k^2 -\textit{i} k a p_{ss})} \; 
\end{aligned}
\end{equation}
\end{widetext}
where $\lambda_T = \lambda_1 + \lambda_3 - 2 \lambda_2$. In this case the sign of the real part of the modes was examined  numerically. We find an instability close to the mean-field transition line in a range of wavevectors along the direction of broken symmetry, analogous to the banding instability of Toner-Tu models~\cite{Bertin2009,Mishra2010}. Near the mean field transition, the banding instability occurs in a narrow region of $s_0$ for $ \sigma \nu>2 |\alpha| \mathrm{D}_p>0 $ and is absent when $\sigma \nu < 0$. A numerical solution of the nonlinear equations reveals, however, a narrow region of banding instability even for $\sigma \nu < 0$. The sign of the morphotaxis parameter $\sigma\nu$ additionally affects the morphology of these banded states (see Fig.~(\ref{fig:phases_sims_combined}C,\ref{fig:phases_sims_combined}D).

Next we examine the stability of the ordered state deep in the flocking regime. In this case fluctuations in the magnitude of polarization,  $\delta p^x_{\mathbf{k}}$, decay on microscopic time scales and can be eliminated by neglecting $\partial_t \delta p^x_{\mathbf{k}} $ in Eq.~(\ref{eq:Mmatrix}). We then obtain coupled equations for fluctuations in cell shape and direction of orientational order. The latter are long-lived at long wavelength because they represent the Goldstone mode associated with the spontaneously broken orientational symmetry. 
The full decay rates are shown in Appendix \ref{app:flocking_stab}. We examine the stability  by carrying out a small wavevector expansion of the hydrodynamic modes.
For $\mathbf{k}=k \mathbf{\hat{x}}$,  corresponding to bend deformation, the homogeneous state is always stable.  
For $\mathbf{k} = k\mathbf{\hat{y}}$,  coupled  splay and shape fluctuations become unstable for
%
\begin{equation}\label{eq:hydrostability}
\sigma \nu  > \frac{ \sigma \lambda_2 a}{\beta_p} + 2 |\alpha| \left( \mathrm{D}_p - \frac{\lambda_2 \lambda_3}{\beta_p} \right)\;.
\end{equation}
%
Unlike the corresponding instability obtained in the Toner-Tu model ~\cite{Mishra2010}, this instability persists even when the advective nonlinearities proportional to $\lambda_2$ and $\lambda_3$ are neglected.

\subsection{\label{subsec:numerics} Numerical simulations}
We have solved numerically the full nonlinear hydrodynamic equations (Eqs.~(\ref{eq:mdot},\ref{eq:pdot})) on  a periodic grid using a standard RK4 explicit iterative method. We choose a timestep $\Delta t = 0.005$ and grid spacing $\Delta x = 0.1$ to satisfy the Von Neumann stability condition. Simulations are initialized in the appropriate homogeneous state (Table.(\ref{tab:steadystate}))  with superimposed spatially white noise of variance small compared to all equation parameters. 
To quantify the onset of spatial patterns, we examine the Fourier spectrum of the configurations obtained at  long times.
If the integral of the discrete Fourier transform of the deviations of the $m$-field from its mean value is greater than some small cutoff number, then the corresponding state is identified as patterned in Fig.(\ref{fig:phases_sims_combined}). Because the perturbations are small, we expect these numerics to agree with and reinforce our analytic phase diagram.

As shown in Fig.~(\ref{fig:phases_sims_combined}) the numerical results agree well with those of the linear stability analysis. For $\nu<0$ (Fig.~(\ref{fig:phases_sims_combined}) top left) the homogeneous states are stable in most of parameter space, with patterns emerging only in a narrow banding region. In contrast, for $\nu>0$ (Fig.~(\ref{fig:phases_sims_combined}) bottom left) we obtain a variety of emergent patterns, as expected from the linear stability analysis. 
As anticipated in Sec.~\ref{subsec:linstab}, the stability boundary of the $\nu=1$ homogeneous solid is shifted as compared to the analytic prediction (i.e. there are blue circles denoting numerical observations of homogeneous states in the region linear stability analysis suggests should be unstable). This is due to the  $m>0$ restriction used in the numerics but not in the linear analysis, which prevents  some instabilities from arising. Reassuringly, we find that relaxing this constraint in simulations resolves the discrepancy and yields agreement with the analytics.



The simulations also  reveal the structure of the spatial patterns that replace the uniform states. Examples are shown in Fig.~(\ref{fig:phases_sims_combined}). For $\nu=1$, in the solid phase we find  droplets of fluid asters surrounded by solid tissue with a positively splayed polarization field (frame A). As $s_0$ increases, the asters become more closely spaced\mdc{, and elongated inclusions begin to appear}. Past the transition from the solid into the liquid, \mdc{these patterns invert} and we find clusters 
of solid tissue surrounded by fluid, with the polarization now pointing inward, corresponding to negative splay (frame B).  In the banding region we observe elongated regions of fluid tissue,  with outward pointing polarization (frame C). Because of the symmetry of the polarization in these bands, the structures do not migrate and their dynamics is reminiscent of coalescence.  The banding patterns obtained for $\nu=-1$ are qualitatively different, as shown in Fig.~(\ref{fig:phases_sims_combined}D). In this case we obtain alternating solid/fluid \mcm{traveling} bands with the polarization aligned transverse to bands. 
The direction of motion of the band is opposite to that direction of the net polarization, which is reminiscent of  a``traffic wave'' phenomenon. 
\section{Conclusions}
We have developed a hydrodynamic theory of confluent tissue close to the recently proposed rigidity  transition tuned by cell shape~\cite{Bi2014,Bi2015,Bi2016}. The hydrodynamic equations are formulated in terms of a scalar field that quantifies single-cell anisotropy and a cell polarization field. Cell anisotropy can drive alignment of local polarization, resulting in a flocking liquid state. The interplay of cell shape and polarization additionally drives the organization of   a variety of aster and banding patterns consisting of solid tissue inclusions in a liquid matrix  or liquid inclusions in the solid, with associated polarization patterns. Pattern selection is controlled by a single parameter $\nu\sigma$, referred to as the morphotaxis parameter, that quantifies the tendency of cells to move towards more rigid or less rigid regions of the tissue.

Since cell anisotropy is effectively a measure of the rheological properties of the tissue, with isotropic cell shapes identifying the solid or jammed state and anisotropic shapes corresponding to a liquid, variations in cell shape anisotropy are directly associated with variation in the rheological properties of the tissue. Our work therefore quantifies for the first time the role of  gradients in  tissue stiffness in driving morphological patterns. This is achieved through a morphotaxis parameter  that couples polarization to gradients of cell shape anisotropy.  Tambe \emph{et al.}~\cite{Tambe2011} used the name ``plithotaxis'' to describe the observed tendency of cells to move in the direction that minimizes local shear stresses. The parameter $\nu$ in our equations could be related to such a plithotactic effect as it embodies the trasmission of  positional sensing in collective cell migrations  via gradient in local tissue rigidity arising from variations in cell shape (see the term $\nu\bm\nabla m$ in Eq.~\eqref{eq:pdot}). Patterning in our model is controlled, however, by the combined action of this term and the changes in local cell shapes induced by polarization sinks and sources (the term $\sigma\bm\nabla\cdot\mathbf{p}$ in Eq.~\eqref{eq:mdot}). These two effects together define the ``morphotaxis'' properties of the tissue.  Our work therefore provides a complementary, purely mechanical view to how patterns of growth and differentiation may be specified in development and tissue regeneration.  Our results could be tested in experiments by combining segmented cell images with traction force microscopy and particle image velocimetry. In solid regions, where cell migration is strongly suppressed, traction forces provide  a direct measure of local cell polarization. Correlating traction measurements with cell shapes could therefore provide information on the sign of the morphotaxis parameter.

Once elongated, cells can also align their orientations and exhibit nematic order on tissue scales, an effect not included in our work. Nematic order has been observed for instance in mouse fibroblasts and can be enhanced by confinement~\cite{Duclos2014}. Recent work has also established an intriguing connection between topological defects in nematic tissue and cell extrusion and death~\cite{Saw2017,Kawaguchi2017}. Work concurrent to ours by Ishihara \textit{et. al.}~\cite{Ishihara2017} has examined the interplay of nematic alignment of elongated cells with tissue mechanical properties and active contraction-elongation. This is accomplished with a continuum model that, although similar in spirit to ours, does not highlight the important distinction between cell anisotropy and nematic order that allows for the onset of polarized states even in the absence of nematic alignment of cell shape, as seen in simulations of self-propelled Voronoi models. Further work will be needed to examine the interplay between cell shape, polarization and nematic order, as well as the role of cell growth, in driving tissue patterning.

\begin{acknowledgements}
We thank Tom Lubensky for useful discussions. We acknowledge support from: the Simons Foundation Targeted Grant in the Mathematical Modeling of Living Systems 342354 (MCM); Simons Foundation grants 446222 and 454947 (MLM); the Syracuse Soft \& Living Matter Program (MC, MLM and MCM); the National Science Foundation DMR-1609208 (MCM), DGE-1068780 (MC and MCM) and DMR-1352184 (MLM); the National Institute of Health R01GM117598-02 (MLM).
\end{acknowledgements}


\appendix
\section{Anisotropic Perturbation of the Shape Tensor}\label{app:shaperturb}
We describe here two  ways in which the shape energy of an irregular polygon may be obtained  as a  perturbation of that of a regular  one. In this section we work at the single-cell level and for convenience suppress the cell label $a$. Using the definition given in  Eq.~\eqref{eq:shapetensor}, the shape tensor $\mathbf{G}^{reg} $ of a regular polygon 
is diagonal and has a single eigenvalue $\lambda$, i.e., it takes the form
\begin{equation}\label{eq:shapetensordiagapp}
\mathbf{G}^{reg} =\lambda \begin{bmatrix} 1 & 0 \\ 0 & 1 \end{bmatrix}= \frac{\mathrm{Tr}\mathbf{G}^{reg}}{2}  \begin{bmatrix} 1 & 0 \\ 0 & 1 \end{bmatrix} \, .
\end{equation}
We are interested in the form that the tensor takes when perturbed away from this initial reference state. As we will only be concerned with quantities constructed from the eigenvalues of this tensor, we may choose to consider the perturbed tensor in a reference frame in which it is diagonal. The perturbed shape tensor can then be written as
%
%
%
\begin{equation}\label{eq:shapetensordiag}
\mathbf{G}^{diag} = \frac{\mathrm{Tr}\mathbf{G}^{reg} + \Delta}{2} \left(  \begin{bmatrix} 1 & 0 \\ 0 & 1 \end{bmatrix} + m \begin{bmatrix} 1 & 0 \\ 0 & -1 \end{bmatrix}\right)\;,
\end{equation}
where $\Delta$ is the change in the tensor trace due to the perturbation. Our choice of the function $\Delta$ will constrain our perturbation to a subset of possible trajectories. Our goal is to show that the choice of this function (within reasonable bounds) is not consequential, and therefore that we may consider the energy in terms of the anisotropy $m$ alone. Employing the Area and Perimeter relations (Eqs.(\ref{eq:shapearea},\ref{eq:shapeperi})), the dimensionless vertex model energy for a single cell can be rewritten in terms of $\Delta$ and $m$ as
\begin{equation}
\begin{aligned}\label{eq:vertexenergyapp}
\epsilon=& \left[\frac{n}{2}\sin(2 \pi/n)(1-m^2)\left(\mathrm{\tilde{Tr}}[\mathbf{G}^{reg}] + \tilde{\Delta}(m)\right)^2-1\right]^2 \\
& +\bar{\kappa} \left[2 n \sin(\pi/n)\left(\mathrm{\tilde{Tr}}[\mathbf{G}^{reg}] + \tilde{\Delta}(m) \right)-s_0\right]^2 \;,
\end{aligned}
\end{equation}
where $\mathrm{\tilde{Tr}}[\mathbf{G}^{reg}] = \mathrm{Tr}[\mathbf{G}^{reg}]/\sqrt{A_0}$ and $\tilde{\Delta}(m) = \Delta(m)/\sqrt{A_0}$ are  dimensionless quantities.

%
%
\begin{figure}[!t]  
    \includegraphics[width=0.4\textwidth]{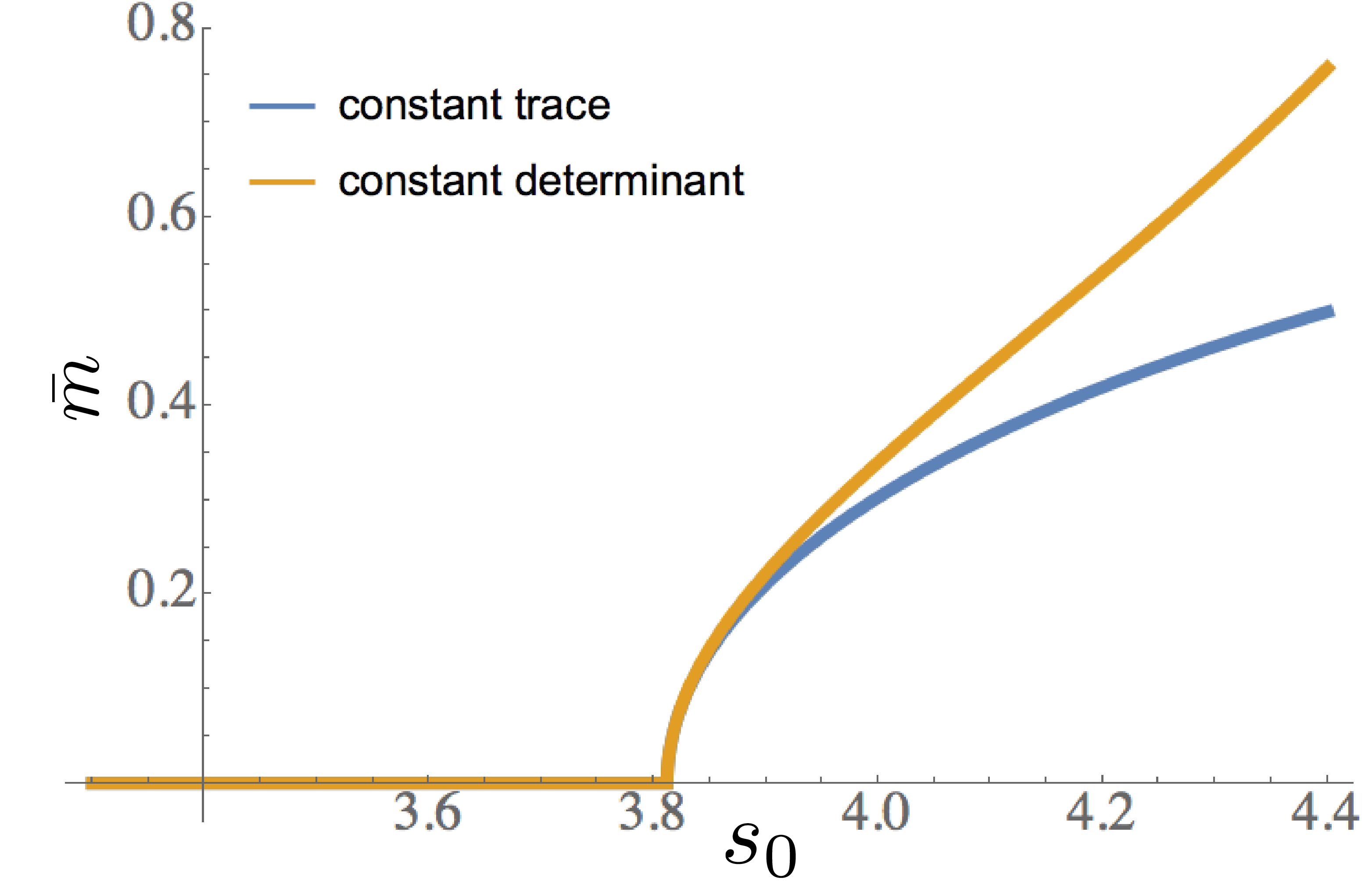}
    \caption{\label{fig:comparemstar} \mcm{The mean value $\bar{m}$ of the  order-parameter obtained by minimizing the single-cell free energy derived using different geometric perturbations of the energy of a regular pentagon.} }
\end{figure}

We first explore the choice $\tilde{\Delta}(m) = 0$ that corresponds to a perturbation with constant trace, hence constant perimeter.
In this case the cell energy becomes
\begin{equation}\label{eq:freeenergyPapp}
\epsilon = \frac{1}{2}\alpha^{(tr)}(\mathrm{\tilde{Tr}} \mathbf{G}^{reg} ) m^2 + \frac{1}{4}\beta^{(tr)}(\mathrm{\tilde{Tr}}\mathbf{G}^{reg}) m^4\;,
\end{equation}
with
\begin{equation}\label{eq:alphatraceapp}
\begin{split}
\alpha^{(tr)}(\mathrm{\tilde{Tr}} \mathbf{G}^{reg} ) 
= &
 2 n \sin(2 \pi / n) (\mathrm{\tilde{Tr}} \mathbf{G}^{reg})^2 \\
&
 -n^2 \sin^2(2 \pi / n) (\mathrm{\tilde{Tr}} \mathbf{G}^{reg})^4\;,
\end{split}
\end{equation} 
\begin{equation}\label{eq:betatraceapp}
\beta^{(tr)}(\mathrm{\tilde{Tr}} \mathbf{G}^{reg} )  = n^2 \sin^2(2 \pi / n) (\mathrm{\tilde{Tr}} \mathbf{G}^{reg})^4 \;,
\end{equation}
where we have shifted the energy by an overall constant, independent of $m$. \mdc{Eqs.~\eqref{eq:alphatrace} \& \eqref{eq:betatrace} may now be recovered from the above by setting $\mathrm{\tilde{Tr}}\mathbf{G}^{reg} = s_0 (2 n \sin{\pi/n})^{-1}$, or equivalently $P=P_0$.}

An alternative approach consists of perturbing $\mathbf{G}^{reg} $ while keeping its determinant constant, implying constant area. The $\tilde{\Delta}(m)$ that preserves this condition is given by
\begin{equation}\label{eq:deltadeterminantapp}
\tilde{\Delta}(m) = \mathrm{\tilde{Tr}}[\mathbf{G}^{reg}] \left( \frac{1}{\sqrt{1-m^2}} -1 \right) \; .
\end{equation}
Using this, the single-cell energy may be written in terms of $m$ and the fixed (dimensionless) area $\tilde{A}$. This energy has the same form as given in Eq.~(\ref{eq:freeenergyPapp}), but with coefficients now given by
%
\begin{equation}\label{eq:alphadet}
\alpha^{(det)}(s_0) =  \tilde{\kappa} \tilde{A} \left[ 8  n \tan{\frac{\pi}{n}} - 4 \frac{s_0}{\sqrt{\tilde{A}}} \sqrt{n \tan{\frac{\pi}{n}}} \right]\;,
\end{equation}
and
\begin{equation}\label{eq:betadet}
\beta^{(det)}(s_0) =  \tilde{\kappa} \tilde{A} \left[ 16  n \tan{\frac{\pi}{n}} - 6 \frac{s_0}{\sqrt{\tilde{A}}} \sqrt{n \tan{\frac{\pi}{n}}} \right] \;.
\end{equation}
Because this energy corresponds to a free cell, the fixed area is expected to realize the target area which implies $\tilde{A}=1$.

\mcm{The value $\bar{m}$ of $m$ that minimizes the single-cell energy (\ref{eq:freeenergyPapp}) for $\alpha<0$  is  $\bar{m}= \sqrt{-\alpha/\beta}$, where $\alpha$ and $\beta$ are given by Eqs~(\ref{eq:alphatrace},\ref{eq:betatrace}) or by Eqs.~(\ref{eq:alphadet},\ref{eq:betadet}) for each of the two perturbations used. The dependence of $\bar{m}$ on $s_0$ for pentagonal cells ($n=5$) obtained using the two perturbations shown in Fig.~(\ref{fig:comparemstar}) demonstrates that the behavior does not depend on the perturbation near the transition, which is  the region of interest in our work. In the main text we use the results obtained with the perturbation that keeps the trace constant.} 
%
%
%
%

\section{\label{app:linstab}Linear Stability Analysis}
The stability analysis follows a  well-known procedure. We consider the equations
\begin{equation} 
\begin{aligned}\label{eq:mdotapp}
\partial_t{m}+ &\nu_1\mathbf{p}\cdot\mathbf{\nabla}m + \nu_2 m \mathbf{\nabla}\cdot\mathbf{p} =-\left[\alpha(s_0)+\beta(s_0) m^2\right]m \\ &+ \sigma\mathbf{\nabla}\cdot\mathbf{p}+\mathrm{D}\nabla^2m\;,
\end{aligned}
\end{equation}
and 
\begin{equation} 
\begin{aligned}\label{eq:pdotapp}
\partial_t{\mathbf{p}}+&\lambda_1\left(\mathbf{p}\cdot\mathbf{\nabla}\right)\mathbf{p}=-\left[\alpha_p - a m + \beta_p p^2\right]\mathbf{p}\\
&-\nu\mathbf{\nabla}m+\lambda_2\mathbf{\nabla}p^2-\lambda_3\left(\mathbf{\nabla}\cdot\mathbf{p}\right)\mathbf{p}+\mathrm{D}_p\nabla^2\mathbf{p} \; ,
\end{aligned}
\end{equation}
where we have included the $\nu_2$ term for generality. To recover the results of the main text, one needs only to set $\nu_2=0$ in  the following equations. Equations (\ref{eq:mdotapp},\ref{eq:pdotapp}) have the uniform, steady state solutions $(m_{ss}, \mathbf{p}_{ss})$  enumerated in Table.(\ref{tab:steadystate}). There are two types of solutions: stationary or non-polarized ones with $|\mathbf{p}_{ss}|=0$ (a fluid and a solid) and moving or polarized ones with  $|\mathbf{p}_{ss}|\not=0$ (flocking fluid). To evaluate the stability of these steady states, we perturb the steady state  solutions ($m \rightarrow m_{ss} + \delta m$,$\mathbf{p} \rightarrow \mathbf{p}_{ss} + \delta \mathbf{p}$) and examine the linear dynamics of the fluctuations (\ref{eq:mdotapp},\ref{eq:pdotapp}). By introducing  Fourier transforms, the linear equations for the fluctuations can be written as
\begin{equation}\label{eq:matrixeqn}
\partial_t\left[ \begin{array}{c}  m_{\mathbf{k}}(t) \\ \rule[-0.5ex]{0pt}{3.5ex} p^x_{\mathbf{k}}(t) \\ \rule[-0.5ex]{0pt}{3.5ex}  p^y_{\mathbf{k}}(t) \end{array} \right] = \mathrm{\textbf{M}}(\mathbf{k}) \left[ \begin{array}{c} m_{\mathbf{k}}(t) \\ \rule[-0.5ex]{0pt}{3.5ex}  p^x_{\mathbf{k}}(t) \\ \rule[-0.5ex]{0pt}{3.5ex} p^y_{\mathbf{k}}(t) \end{array} \right]\;,
\end{equation}
where 
\begin{equation}\label{eq:fourierperturbapp}
\left[ \begin{array}{c}  m_{\mathbf{k}}(t) \\ \rule[-0.5ex]{0pt}{3.5ex} p^x_{\mathbf{k}}(t) \\ \rule[-0.5ex]{0pt}{3.5ex}  p^y_{\mathbf{k}}(t) \end{array} \right] = \int \frac{\mathrm{d}\mathbf{x}}{(2 \pi)^2} ~e^{i \mathbf{k} \cdot \mathbf{x}}\left[ \begin{array}{c} \delta m(\mathbf{x},t) \\ \rule[-0.5ex]{0pt}{3.5ex} \delta p^x(\mathbf{x},t) \\ \rule[-0.5ex]{0pt}{3.5ex}  \delta p^y(\mathbf{x},t) \end{array} \right] 
\end{equation}
are the Fourier amplitudes and the explicit expression of the matrix $\mathrm{\textbf{M}}(\mathbf{k})$ depends on the homogenenous state considered. We seek solutions of the form
\begin{equation}
( m_{\mathbf{k}}(t), p^x_{\mathbf{k}}(t) , p^y_{\mathbf{k}}(t) ) = \mathrm{exp}(z t)( m_{\mathbf{k}}, p^x_{\mathbf{k}} , p^y_{\mathbf{k}} ) \; .
\end{equation}
The eigenvalues of $\mathrm{\textbf{M}}(\mathbf{k})$ then represent the growth rates of the perturbations.

A homogeneous state is then linearly stable iff the real part of each eigenvalue of $\mathrm{\textbf{M}}(\mathbf{k})$ is negative for all $\mathbf{k}$. With this condition satisfied, all small perturbations   decay in time and the system returns to the steady state.  
The lack of symmetry breaking in the non-polarized regimes allows $\mathrm{\textbf{M}}(\mathbf{k})$ and the stability analysis to be simplified greatly. We consider these solutions first. 
\subsection{Stability of stationary (non-polarized ) states}
First, we analyze the region in which $m_{ss}=|p_{ss}|=0$. Here,  $\mathrm{\textbf{M}}(\mathbf{k})$ is simplified by considering $\mathbf{p}_{\mathbf{k}} = p^L_{\mathbf{k}} \mathbf{\hat{k}} + p^T_{\mathbf{k}} \mathbf{\hat{k}}_{\perp}$ as shown in Eq.~\eqref{eq:matrix1}. This form, for later convenience, applies to both the fluid and solid.
\begin{widetext}
\begin{equation}\label{eq:matrix1}
\mathrm{\textbf{M}}^{iso}(\mathbf{k})=\\ \left[ \begin{array}{c c c} -\alpha -3 \beta m_{ss}^2-\mathrm{D} k^2 & -i\sigma(m_{ss})k & 0 \\ 
i \nu k & -\alpha_p(m_{ss}) - \mathrm{D}_p k^2 & 0 \\ 
0 & 0 & -\alpha_p(m_{ss}) - \mathrm{D}_p k^2 
\end{array} \right]
\end{equation}
\end{widetext}
Where $\alpha_p(m_{ss}) \equiv \alpha_p^0 - a m_{ss}$ and $\sigma(m_{ss}) \equiv \sigma - \nu_2 m_{ss}$. \mcm{We see that fluctuations $p^T_{\mathbf{k}}$ in the transverse polarization, describing bend deformations, are decoupled and always stable, and decay at the rate $z_{\perp} = -\alpha_p(m_{ss}) - \mathrm{D}_p k^2$. }
The other two eigenvalues control  coupled fluctuations in shape and longitudinal polarization $p^L_{\mathbf{k}}$, corresponding to splay deformations and are given by the solutions of a quadratic equation,
\begin{equation} 
\begin{aligned}\label{eq:eigenvals1app}
&2 z_{\pm}(k)=-[\alpha(m_{ss})+\alpha_p(m_{ss})]-(\mathrm{D}+\mathrm{D}_p)k^2 \\ 
& \pm \sqrt{\left[\alpha(m_{ss})-\alpha_p(m_{ss})+(\mathrm{D}-\mathrm{D}_p)k^2\right]^2+4 k^2 \nu \sigma(m_{ss})} \; ,
\end{aligned}
\end{equation}
where $\alpha(m_{ss}) = \alpha + 3 \beta m_{ss}^2$. The stability is always controlled by the mode $z_+(k)$.

\subsection{Stability of Stationary Solid}

\begin{figure}[!t]
\begin{centering}
    \includegraphics[width=0.4\textwidth]{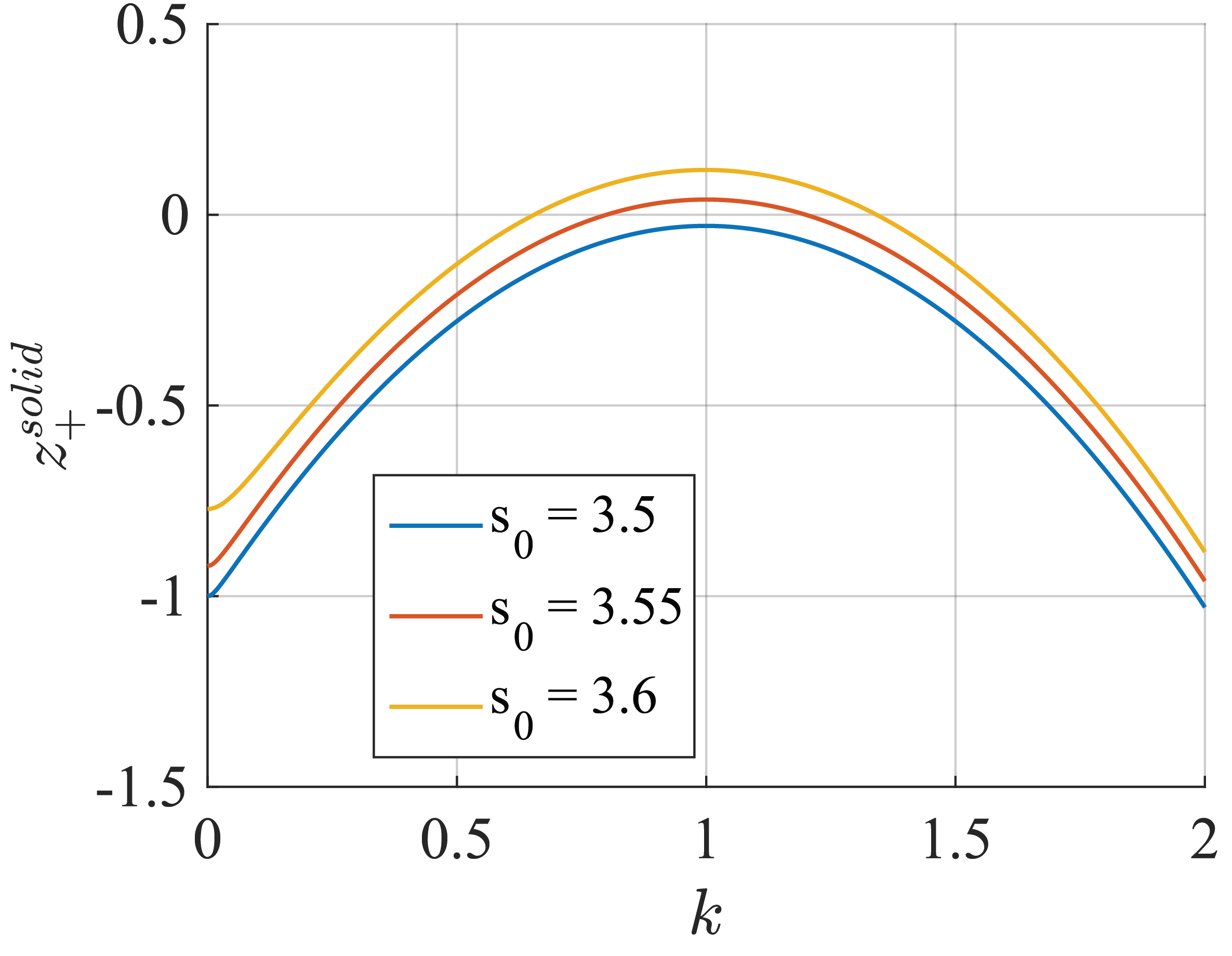}
    \caption{ Dispersion relation of the eigenvalue  $z_+^{solid}(k)$ in the stationary solid phase as a function of $k$. Lines correspond to points in the $\nu=1$ phase diagram from Fig.\ref{fig:phases_sims_combined}.  The blue curve then corresponds to a stable homogeneous state, while yellow and green represent unstable states.}
    \label{fig:mode_example1}
    \end{centering}
\end{figure}

In the solid $m_{ss}=0$, hence $\alpha_p(m_{ss})=\alpha_p^0$ and $\sigma(m_{ss})=\sigma$. Instabilities in the homogeneous stationary solid will arise $(z_+(k)>0)$ when 
\begin{equation}\label{eq:stability1}
\sigma \nu > \left(\sqrt{\alpha_p^0\mathrm{D}} + \sqrt{\alpha\mathrm{D}_p} \right)^2 
\end{equation}
in a band of wavectors $k_-<k<k_+$. The wavevectors $k_\pm$ are solutions of a quadratic equation 
\begin{equation}\label{eq:proxystability1}
\alpha \alpha_p^0 + \left[ \alpha \mathrm{D}_p + \alpha_p^0 \mathrm{D} - \sigma \nu \right] k^2 + \mathrm{D} \mathrm{D}_p k^4 = 0
\end{equation}
and are given by
\begin{equation}
\begin{aligned}\label{eq:unstablerange1}
k_{\pm}^2 &= - \left[\frac{\alpha}{2\mathrm{D}} + \frac{\alpha_p^0}{2 \mathrm{D}_p} - \frac{\sigma \nu}{2\mathrm{D}\mathrm{D}_p} \right] \\ 
& \pm \sqrt{\left[\frac{\alpha}{2\mathrm{D}} + \frac{\alpha_p^0}{2 \mathrm{D}_p} - \frac{\sigma \nu}{2\mathrm{D}\mathrm{D}_p}\right]^2 - \frac{\alpha \alpha_p^0}{\mathrm{D} \mathrm{D}_p} } \; .
\end{aligned}
\end{equation}
These solutions are real provided Eq.~\eqref{eq:stability1} is satisfied.
The dispersion relation of the mode $z_+(k)$ in the stationary solid phase is shown in Fig.~(\ref{fig:mode_example1}) for a few parameter values. 
Note that an instability can only occur provided $\sigma\nu > 0$.
Near the onset of instability the wavelength of the fastest growing mode is given by
%
\begin{equation}\label{eq:lengthscale1app}
\ell^{solid} =\left (\frac{\mathrm{D}\mathrm{D}_p}{\alpha\alpha_p^0}\right)^{1/4} \; ,
\end{equation}
which is the geometric average of the length scale $l_m(m_{ss})=\sqrt{\mathrm{D}/\alpha(m_{ss})}$ governing variation in $m$ and the length scale $l_p(m_{ss})=\sqrt{\mathrm{D}_p/\alpha_p(m_{ss})}$ controlling spatial variation of the polarization $\mathbf{p}$. The lengths $l_m$ and $l_p$ represent the characteristic distances over which diffusion balances the decay rate. In the solid phase, we find emergent patterns as $s_0$ is increased. Further increase of $s_0$ increases the characteristic scales of such patterns. Because  $\alpha(s_0)$ is roughly linear in $s_0$ in the range of interest, we may deduce the critical scaling  as $l \sim (s_0-s_0^*)^{0.25}$, where  $s_0^*=3.812$.

\subsection{Stability of Stationary Fluid   }
In the fluid, we  have $m_{ss}^2 = -\alpha /\beta$, $\alpha_p(m_{ss}) = \alpha_p^0 - a m_{ss}$ and $\sigma(m_{ss}) = \sigma  - \nu_2 m_{ss}$. When
\begin{equation}\label{eq:stability2}
 \nu \sigma(m_{ss}) > \left(\sqrt{\alpha_p(m_{ss})\mathrm{D}} + \sqrt{2|\alpha|\mathrm{D}_p} \right)^2
\end{equation}
the mode $z_+$ is unstable for a band of wavevectors $k_- < k < k_+$. The wavevectors $k_\pm$ are again solutions of a quadratic equation
\begin{equation}\label{eq:proxystability2}
\alpha \alpha_p(m_{ss}) + \left[ \alpha \mathrm{D}_p + \alpha_p(m_{ss}) \mathrm{D} - \sigma(m_{ss}) \nu \right] k^2 + \mathrm{D} \mathrm{D}_p k^4 = 0
\end{equation}
and are given by
\begin{equation}
\begin{aligned}\label{eq:unstablerange2}
&  k_{\pm}^2 = - \left[\frac{\alpha_p(m_{ss})}{2\mathrm{D}_p} - \frac{\alpha}{\mathrm{D}}  -  \frac{\nu \sigma(m_{ss})}{2 \mathrm{D} \mathrm{D}_p} \right] \\
&\pm \sqrt{\left[\frac{\alpha_p(m_{ss})}{2\mathrm{D}_p} -\frac{\alpha}{\mathrm{D}} - \frac{\nu \sigma(m_{ss})}{2 \mathrm{D} \mathrm{D}_p} \right]^2 + 2 \frac{ \alpha \alpha_p(m_{ss})}{\mathrm{D} \mathrm{D}_p} } \; .
\end{aligned}
\end{equation}
From this equation we are able to isolate the stability condition 
as well as the characteristic wavevector of the fastest growing mode near the stability-instability boundary. This gives us a lengthscale 
\begin{equation}\label{eq:lengthscale2}
\ell^{fluid}= \sqrt{-\mathrm{D} \mathrm{D}_p / 2 \alpha (\alpha_p^0 -a m_{ss})}
\end{equation}
expected to govern emerging patterns. Again this may be thought of as the geometric average of the length scales $\ell_m(m_{ss})$ and $\ell_p(m_{ss})$ controlling spatial variation in the decoupled fields.

\subsection{\label{app:flocking_stab} Stability of Flocking Fluid} 

In this case there is special direction in the system, which is the direction of the broken-symmetry $\mathbf{p}_{ss} \neq 0 $, and all modes are coupled. 
The stability matrix is given by
\begin{widetext}
\begin{equation}\label{eq:matrix3}
\mathrm{\textbf{M}}^{pol}(\mathbf{k})=
\left[ \begin{array}{c c c} 2\alpha  + i\nu_1 p_{ss} k_x-\mathrm{D} k^2 & -i\sigma(m_{ss})k_x & -i\sigma(m_{ss}) k_y \\ 
i \nu k_x + a p_{ss} & 2\alpha_p(m_{ss}) + i\lambda_{T}p_{ss} k_x - \mathrm{D}_p k^2 & i \lambda_3 p_{ss} k_y \\ 
i \nu k_y & -2 i \lambda_2 p_{ss} k_y & i\lambda_1 p_{ss} k_x - \mathrm{D}_p k^2 
\end{array} \right] \; ,
\end{equation}
\end{widetext}
where \mcm{we have chosen a coordinates system with the $x$ axis along the direction of broken symmetry,} so that $\mathbf{p}_{ss} = p_{ss}\mathbf{\hat{x}}$. We have  defined $\lambda_{T} = \lambda_1 - 2\lambda_2 + \lambda_3$. To avoid solving a cubic equation for the decay rates, we only estimate stability along special directions.

\subsubsection{Banding Instability}
\mcm{We first  examine the behavior of the modes for $\mathbf{k}$ along the direction of broken symmetry, $\mathbf{k}=k\mathbf{\hat{x}}$. Fluctuations in $p_{\mathbf{k}}^y$ then decouple and are always stable. }
\mcm{The quadratic equation for the remaining two modes is easily solved, with the result}
%
\begin{widetext}
\begin{equation}
\begin{aligned}\label{eq:eigenvalbanding}
2 z^{(band)}_{\pm}= & 2(\alpha +\alpha_p(m_{ss})) +\textit{i} p_{ss} (\nu_1 + \lambda_{T}) k - (\mathrm{D} +\mathrm{D}_p) k^2 \\ 
& \pm \sqrt{\left[2(\alpha-\alpha_p(m_{ss}))+\textit{i} p_{ss} (\nu_1-\lambda_{TOT}) k - (\mathrm{D}-\mathrm{D}_p) k^2\right]^2+4\sigma(m_{ss})(\nu k^2 -\textit{i} k a p_{ss})} \; .
\end{aligned}
\end{equation}
\end{widetext}
\mcm{Close to the mean-field transition between stationary and flocking liquid ($\alpha_p(m_{ss})=0$)  a small wavevector expansion yields an instability for}
%
\begin{equation}\label{eq:bandingcondition}
  \nu \sigma(m_{ss}) > 2 |\alpha| \mathrm{D}_p>0 \, .
\end{equation}
\mcm{The instability boundaries shown in our phase diagram are obtained, however, through a more general analysis carried out with Mathematica.
The wavelength of the fastest growing mode can also be calculated. In the limit 
$\alpha_p(m_{ss})\rightarrow 0$ it is given by}
\begin{equation}\label{eq:bandinglength}
\ell_{band} \sim  \frac{\pi}{|\alpha|} \sqrt{ \frac{2 \sigma(m_{ss}) \nu |\alpha|(\mathrm{D}_p-\mathrm{D}) - \sigma(m_{ss})^2 \nu^2}{|\alpha|\mathrm{D}_p - \frac{1}{2} \sigma(m_{ss}) \nu}} \, .
\end{equation}
\mcm{This instability is analogous to the banding instability of Toner-Tu models, as it describes the onset of bands of alternating ordered and disordered regions preferentially aligned in the direction transverse to that of broken symmetry.}

\subsubsection{\mcm{Instability of Splay Fluctuations}}
We now analyze the stability deep in the ordered polar state. 
In this region, fluctuations in $p_{ss}$ \mcm{always decay on short time scales.} 
\mcm{For this reason we neglect  $\partial_t p_x $ and eliminate $p_x$ in favor of $p_y$ and $m$, obtaining again a quadratic equation for the dispersion relation of the modes that can be solved analytically.  For simplicity we only examine the modes for} $\vec{k} = k_x \hat{x}$ and $\vec{k} = k_y \hat{y}$.  These  decay rates of the hydrodynamic mode are then given by 
\begin{equation}\label{eq:eigenval3x}
 z^{(flock)}_{+}(k_x)=\textit{i} A_x k_x-\mcm{D_p} k_x^2 + \mathcal{O}(k_x^3) 
\end{equation}
and 
\begin{equation}\label{eq:eigenval3y}
 z^{(flock)}_{+}(k_y)=\textit{i} A_y k_y-\mathrm{D}^{eff}_y k_y^2 + \mathcal{O}(k_y^3) \; .
\end{equation}
%
\mcm{The mode is always stable for $\vec{k} = k_x \hat{x}$. 
In contrast,}
\begin{equation}\label{eq:hydrodiffusion_y}
\mathrm{D}^{eff}_y=\mathrm{D}_p - \frac{\lambda_2 \lambda_3}{\beta_p} - \frac{\sigma(m_{ss})}{2 |\alpha|}\left(\nu - \frac{a \lambda_2}{\beta_p}\right) 
\end{equation}
\mcm{changes sign, resulting in the coupled instability of shape anisotropy and splay fluctuations of the polarization for}
%
\begin{equation}\label{eq:hydrostabilityapp}
\sigma(m_{ss}) \nu  > \frac{ \sigma(m_{ss}) \lambda_2 a}{\beta_p} + 2 |\alpha| \left( \mathrm{D}_p - \frac{\lambda_2 \lambda_3}{\beta_p} \right) \, .
\end{equation}

\bibliography{bibliofile}
\end{document}